\documentclass[conference]{IEEEtran}
\usepackage{subcaption}
\usepackage[english]{babel}
\usepackage[utf8]{inputenc}
\usepackage{amsmath}
\usepackage{amsfonts}

\usepackage{amsthm}
\usepackage{mathtools}
\usepackage{graphicx}
\graphicspath{{Figures/}}
\usepackage{epstopdf}

\usepackage{epsfig}
\usepackage{url}
\usepackage{comment}
\usepackage{graphics}
\usepackage{soul} 
\usepackage{multirow}

\usepackage{todonotes} % to do (added by monowar)

\usepackage{graphicx, array} % for figure inside table

\usepackage{wrapfig}

\usepackage{algorithm}  
\usepackage{algorithmic}

\usepackage{listings}
\usepackage{xspace}

\usepackage{enumerate}% http://ctan.org/pkg/enumerate
\usepackage{paralist} %inline list
\usepackage{array}
\usepackage{ragged2e}
\newcolumntype{P}[1]{>{\RaggedRight\hspace{0pt}}p{#1}}

\usepackage{relsize}

\usepackage{multirow}

\makeatletter
\renewcommand{\fnum@figure}{Fig. \thefigure}
\makeatother

\usepackage{nimbusmononarrow}
\usepackage{lstlinebgrd}

\definecolor{mygreen}{rgb}{0,0.6,0}
\definecolor{mygray}{rgb}{0.5,0.5,0.5}
\definecolor{mymauve}{rgb}{0.58,0,0.82}
\definecolor{mylightgray}{rgb}{0.7421875,0.7421875,0.7421875}

\colorlet{shadecolor}{gray!20}

\newcommand{\eg}{{\it e.g.,}\xspace}\soulregister{\eg}{7}
\newcommand{\viz}{{\it viz.,}}

\newcommand{\etal}{{\it et~al.}}
\newcommand{\ie}{{\it i.e.,}\xspace}\soulregister{\ie}{7}
\newcommand{\etc}{{\it etc.}}\soulregister{\etc}{7}
\newcommand{\ci}{{\it (i) }}
\newcommand{\cii}{{\it (ii) }}
\newcommand{\ciii}{{\it (iii) }}

\newcommand{\ca}{{\it (a) }}
\newcommand{\cb}{{\it (b) }}

\newcounter{exml}
\newtheorem{example}[exml]{Example}

\newtheorem{proposition}{Proposition}

\soulregister{\cite}{7}
\soulregister{\ref}{7}
\soulregister{\em}{7}

\def\QED{\ensuremath{\boldsymbol{\square}}}
\def\markatright#1{\leavevmode\unskip\nobreak\quad\hspace*{\fill}{#1}}
\def\qed{\markatright{\QED}}
\usepackage{MnSymbol}

\IEEEoverridecommandlockouts

\begin{document}

%\newcommand{\papername}%{\textsf{reOrDer}\xspace}
% \todo[inline]{RTAS 2018 Submission Deadline: Oct 6, 2017 11:59pm (UTC-12) - firm}
% \todo[inline]{10 pages main content + 2 pages appendix + additional pages permitted for the bibliography and acknowledgments only.}
% \setcounter{page}{0}
% \newpage

% \newcommand{\papername}{\xspace{\textsc{REORDER}}\xspace}
\newcommand{\papername}{\xspace{\relsize{-0.999}\textsc{REORDER}}\xspace}
%\newcommand{\papername}{\textsc{reOrDer}}

% \newcommand{\papername}{\xspace{\textsc{reOrDer}}\xspace}

% \title{\papername: Securing Real-Time Systems Using Schedule Obfuscation}
\title{\papername: Securing Dynamic-Priority Real-Time Systems Using Schedule Obfuscation\thanks{\IEEEauthorrefmark{1}These authors contributed equally to this work.}}

\author{
\IEEEauthorblockN{Chien-Ying Chen\IEEEauthorrefmark{1}, Monowar Hasan\IEEEauthorrefmark{1}, AmirEmad Ghassami, Sibin Mohan and Negar Kiyavash} 
\IEEEauthorblockA{%Coordinated Science Laboratory and 
%\IEEEauthorrefmark{2}Department of Computer Science, 
%\IEEEauthorrefmark{3}Department of Electrical and Computer Engineering, 
University of Illinois at Urbana-Champaign, Urbana, USA}
\{cchen140, mhasan11, ghassam2, sibin, kiyavash\}@illinois.edu
%\vspace{-1.5ex}
}
%\IEEEauthorrefmark{2}

% MH: for page numbers
\thispagestyle{plain}
\pagestyle{plain}

% \saythanks	% This is for the author note. Supported by the "abstract" package

\maketitle

% \begin{abstract}
%\todo[inline]{RTAS 2018 Submission Deadline: Oct 6, 2017 11:59pm (UTC-12) - firm}
%\todo[inline]{10 pages main content + 2 pages appendix + additional pages permitted for the bibliography and acknowledgments only.}
%{The main body of each submitted paper is limited to 10 pages of technical content with additional pages permitted for the bibliography and acknowledgments only. Additionally, each submission may include an optional appendix with supplemental material that will be read at the discretion of the program committee; this appendix is limited to two pages (for up to 12 pages total of technical content, i.e., content not including the bibliography and acknowledgements). Details of the submission process can be found on the conference website.}

%The predictable design of real-time systems (RTS) can often aid attackers.

\begin{abstract}
    % \hl{TODO: Change the symbol for Ready Queue (to something like $RQ$.}
    % % \hl{TODO: Blind the source code link (send the link to the chair as online appendix.}
    % \hl{TODO: Merging the simulation and evaluation so that schedule entropy experiments can be considered a part of evaluation.}
    % \hl{TODO: 10 + with additional pages permitted for the bibliography and acknowledgments only.}
	The deterministic (timing) behavior of real-time systems (RTS) can be
	used by adversaries -- say, to launch side-channel attacks or even
	destabilize the system by denying access to critical resources. We
	propose a protocol (named \papername) to \textit{obfuscate} this
	predictable timing behavior of RTS, especially ones designed using
	dynamic-priority scheduling algorithms (\eg~EDF). We also present a
	metric (named ``schedule entropy'') that measures the levels of
	obfuscation introduced into a given real-time system.  The \papername
	protocol was integrated into the standard Linux real-time scheduler and
	evaluated on a realistic embedded platform (Raspberry Pi) running the
	MiBench automotive benchmark workloads. We also demonstrate how designers of RTS can 
	increase the security of their systems and also quantitatively measure
	the impact (both in terms of security and performance) of using this
	protocol.
\end{abstract}

\section{Introduction}
\label{sec::intro}

Systems with real-time properties are often engineered to be very
predictable\cite{wcet_survey}.  This is necessary for their correct operation
and ensuring safety guarantees. Most real-time systems (RTS) are designed to
execute repeating jobs (either periodic or sporadic\footnote{Jobs with bounded inter-arrival times.} ones) that have explicit ``deadline''
requirements. Hence, the \textit{schedule repeats}. Any deviations in timing
behavior, for the real-time schedule, can result in the system becoming unstable --
thus, adversely affecting the \textit{safety} of the system. Adversaries
can take advantage of this inherent determinism by focusing their attacks on
the schedulers in real-time systems \cite{chen2018scheduleak,chen2015scheduleak}. Traditionally, security
has always been an afterthought in the design of RTS but that is changing with
the advent of high-profile attacks (\eg denial-of-service attacks using
Internet-of-Things devices \cite{ddos_iot_camera}, Stuxnet \cite{stuxnet},
BlackEnergy \cite{Ukraine16}, \etc). The increase usage of
commodity-off-the-shelf (COTS) components along with emerging technologies (\eg
IoT) only exacerbates security problems in RTS.

\begin{figure}[t]
 	%\vspace{-1\baselineskip}
     \centering
     \begin{subfigure}[t]{0.49\textwidth}
         \centering
         \includegraphics[width=0.8\columnwidth]{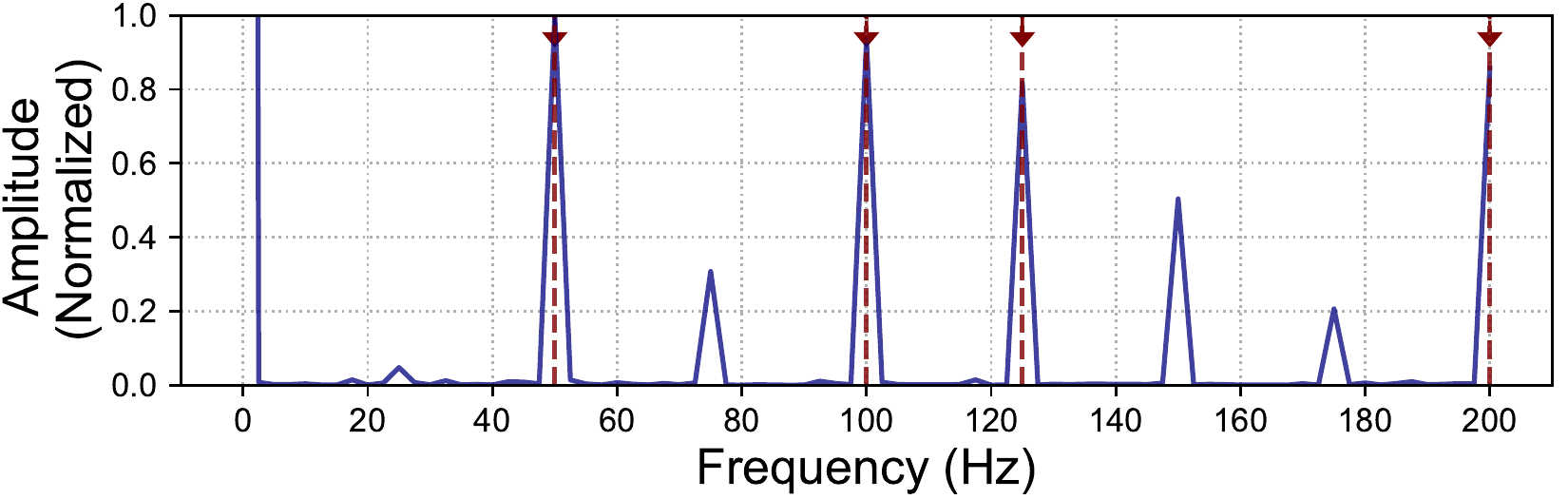}
         \vspace{-0.4\baselineskip}
         \caption{Frequency Spectrum of the EDF Schedule}
 	\label{fig:fft_edf}
         %\vspace{-\baselineskip}
     \end{subfigure}%   
     \hfill  
     \begin{subfigure}[t]{0.49\textwidth}
         \centering
         \includegraphics[width=0.8\columnwidth]{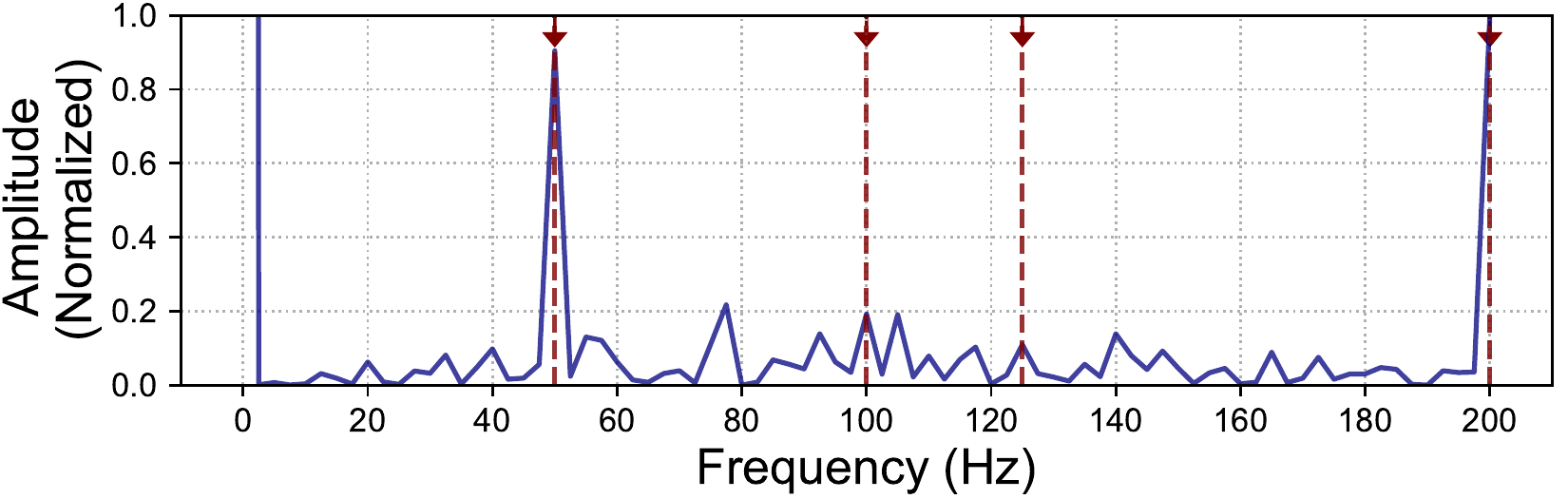}
         \vspace{-0.4\baselineskip}
         \caption{Frequency Spectrum of the Obfuscated Schedule}
         %\vspace{-0.5\baselineskip}
 	\label{fig:fft_reorder}
     \end{subfigure}
     \vspace{-0.5\baselineskip}
     \caption{Results of the frequency spectrum analysis for execution sequences scheduled by: (a) a deterministic scheduler (\eg Vanilla EDF) and (b) randomized scheduler (\papername). A taskset consisting of $4$ periodic tasks 
%([97ms,78ms,64ms,29ms,10ms] as their periods) 
%($[10.3, 12.8, 15.6, 34.5, 100]$ as their frequencies)
is considered and the $4$ arrows (in red) in each plot indicate their true frequencies (\ie $F_i = \tfrac{1}{T_i} = \left\lbrace 50, 100, 125, 200\right\rbrace$ Hz where $T_i$ is the inter-arrival time of the tasks.)}
     \label{fig:fft}
     \vspace{-1.5\baselineskip}
 \end{figure}

Hence, the scheduler in RTS is a critical component for maintaining the
integrity of the system. In fact, the predictable behavior can be used to
improve the security of such systems \cite{securecore,sg1,sg2,mhasan_rtss16}.  On the other hand, there are significant vulnerabilities
that adversaries could exploit due to the repeating nature of the real-time
task schedules. Consider the spectrum analysis of a $4$-task real-time system
(from Example~\ref{ex:simple_taskset} introduced in Section \ref{sec:radom_prio_inv})
using discrete Fourier transforms (DFT) (Fig. \ref{fig:fft_edf}). An
adversary can easily reconstruct the execution frequencies (and hence, periods)
of \textit{all four real-time tasks} (annotated by the red arrows) from this
information! Such information can be used to launch other
attacks\footnote{Attackers can launch these attacks with greater
accuracy/success since they can predict \textit{exactly} when the victim tasks
are released based on the information presented in Fig. \ref{fig:fft_edf}.}
-- \eg scheduler side-channels that can then be used to leak
critical information~\cite{chen2018scheduleak,chen2015scheduleak} or deny critical services, power
consumption~\cite{Jiang2014}, schedule
preemptions~\cite{embeddedsecurity:son2006}, electromagnetic (EM)
emanations~\cite{agrawal2002side} and temperature~\cite{bar2006sorcerer}, \etc~In fact, defensive techniques for RTS is fairly limited
\cite{mohan_s3a,xie2007improving,lin2009static,securecore, sg1, sg2,
sipta,mhasan_rtss16}. 

\textit{Obfuscating the schedules}, \ie introducing randomness into the execution
patterns of real-time tasks, could be one way to improve the security of RTS.
This must be done in a careful manner, so as to not interfere with the timing
guarantees that the system can provide, while still introducing diversity into
the schedule. Figure \ref{fig:fft_reorder} shows the results of applying our
randomization protocol (introduced next) to the same 4-task 
example mentioned earlier. As the figure shows, DFT analysis applied to an 
obfuscated schedule results in less regular execution patterns -- hence, 
it is harder to identify the frequencies of two of the tasks 
(second and third red arrows), thus thwarting some of the potential attacks
mentioned earlier.

% \begin{figure}[t]
% % \vspace{-\baselineskip}
%     \centering
% %     \vspace*{-1em}
% %     \hspace*{-1em}
%     \subfigure[Frequency Spectrum of the EDF Schedule]
%     {
%      %\vspace*{-1\baselineskip}
% %     	\hspace*{-0em}
%         \includegraphics[width=0.8\linewidth]{fft_edf}
%         \label{fig:fft_edf}
%          %\vspace*{-1\baselineskip}
%     }
%     %\\
% %     \hspace*{-1.0em}

%     \subfigure[Frequency Spectrum of the Obfuscated Schedule]
%     {
    
%         \includegraphics[width=0.8\linewidth]{Figures/fft_reorder}

%         \label{fig:fft_reorder}
%     }
%      \vspace{-0.7\baselineskip}
%     \caption{Results of the frequency spectrum analysis for execution sequences scheduled by: (a) a deterministic scheduler (\eg Vanilla EDF) and (b) randomized scheduler (\papername). A taskset consisting of $4$ periodic tasks 
% %([97ms,78ms,64ms,29ms,10ms] as their periods) 
% %($[10.3, 12.8, 15.6, 34.5, 100]$ as their frequencies)
% is considered and the $4$ arrows (in red) in each plot indicate their true frequencies (\ie $F_i = \tfrac{1}{T_i} = \left\lbrace 50, 100, 125, 200\right\rbrace$ Hz).}
%      \label{fig:fft}
%     \vspace*{-1.0\baselineskip}
% \end{figure}

We propose a schedule randomization protocol (Section \ref{sec:protocol_overview}) that we named \papername
(REal-time ObfuscateR for Dynamic SchedulER). We achieve this by using
\textit{bounded priority inversions} at runtime (see Section \ref{sec:radom_prio_inv} for more
details). \papername obfuscates the earliest deadline first (EDF) scheduling
policy~\cite{Liu_n_Layland1973}; EDF is a dynamic task scheduler that can,
theoretically, utilize a CPU to its fullest.  It is widely supported by many
real-world RTS and operating systems, \eg Erika Enterprise \cite{erika}, RTEMS
\cite{rtmes}, \etc~and even Linux \cite{sched_dead}. Existing work on
protecting real-time schedulers~\cite{volp_TT_randomization, taskshuffler}
is \ca focused on static scheduling algorithms and \cb inadequate for
measuring the effects of obfuscation. Obfuscating the schedules for dynamic
priority algorithms such as EDF, to achieve a high level of
randomization, is a much harder proposition than that for static algorithms.
One important problem is how to bound the time allocated for allowing priority
inversions since the job deadlines dynamically change as the execution proceeds\footnote{In static algorithms, these bounds can easily be computed offline and stored in lookup tables.}. 
\papername guarantees that if a given real-time system was schedulable
(\ie meets all of its timing and deadline constraints) by the vanilla EDF scheduler, 
then the \textit{obfuscated schedule will also meet the same guarantees}.

A challenge in any security framework is to \textit{measure} the
effectiveness of the solution. In this case, designers of RTS need to estimate
the amount of randomness introduced into the real-time schedule by \papername. 
Hence, we developed a metric that we named ``schedule entropy'' (Section \ref{subsec::reorder_entropy})
that measures the amount of obfuscation for each given real-time task set/schedule.
Hence, schedule entropy can be used to not just capture the amount of randomness
introduced into the system but also \textit{compare different obfuscation schemes}.

\papername is implemented in a (real-time) Linux kernel\footnote{Please see repository \cite{github:redf} for the source code of our implementation.}. It was evaluated (Section \ref{subsec:linux})
% \papername is implemented in a (real-time) Linux kernel\footnote{Please see \hl{anonymous} repository for the source code of our implementation.}. It was evaluated (Section \ref{subsec:linux})
\ca on a realistic embedded platform (Raspberry Pi) 
\cb using an automotive benchmark suite (MiBench)~\cite{guthaus2001mibench}. In addition, we also carry out a design-space exploration using synthetic real-time
task sets\footnote{A common practice in the real-time community.} (Section \ref{sec:exp_syntehtic}). 
This paper makes the following
contributions:
\begin{itemize}
\item a randomization algorithm that shuffles EDF schedules (Section \ref{subsec:protocol}),
\item ``schedule entropy'' -- a new metric to calculate the amount of randomness
	in the task schedules (Section \ref{sec:entropy}) and
\item an implementation of the \papername algorithm in the Linux kernel (Section 
	\ref{subsec:linux}).
\end{itemize}
We first present some background information as well as the system and adversary models.

\section{System and Adversary Model}

\subsection{Background}

Standard real-time scheduling theory generally considers periodically executing \textit{tasks}\footnote{This trivially maps with the concept of a \textit{process} in general purpose OS.}~\cite{sporadic_task,Liu_n_Layland1973,dbf_sanjay} that models typical real-time control systems. Each task $\tau_i$ generates a potentially infinite sequence of jobs and is modeled  by the worst-case computation time (WCET) $C_i$ and a defined minimum inter-arrival time (\ie period) $T_i$. Also tasks have a strict (relative) deadline $D_i$ by which a computation must be finished. Task priorities can be static or dynamic \cite{Liu_n_Layland1973}. The optimal static scheme is the RM priority assignment where shorter period implies higher priority. RM can guarantee schedulability of a given set of tasks as long as the total utilization is below $\log 2 \approx 69\%$. The overall optimal scheme is EDF -- a dynamic-priority algorithm that always picks the job of a task whose deadline is closest. EDF can schedule any set of tasks if the total system utilization does not exceed 100\% (\eg sum of the WCET to period ratio for all tasks in the system is less than unity: $ \sum\limits_{\tau_i} \tfrac{C_i}{T_i} < 1$).

\subsection{System Model}

%\todo[inline]{TODO:MH fix writeup for more general audience}

Let us consider the problem of scheduling a set of $n$ periodic tasks $\Gamma = \lbrace \tau_1, \tau_2, \cdots ,\tau_n \rbrace $ on a single processor\footnote{Since most RTS are still using single core platforms.}, using the EDF scheduling policy. % In this work we focus on the widely used periodic task model \cite{sporadic_task} in which each task $\tau_i \in \Gamma$ is characterized by the tuple $(C_i, T_i, D_i)$ where $C_i$ is the worst-case execution time (WCET), $T_i$ is the period and $D_i$ is the relative deadline.
%minimum separation (\eg period) between two successive invocations 
%(\ie each task generates a potentially infinite sequence of jobs\footnote{For the simplicity of notation, we use the same symbol $\tau_i$ to denote its jobs and use the term \textit{task} and \textit{job} interchangeably.} with successive job-arrivals separated by at least $T_i$ time units) 
For simplicity of notation, we use the same symbol $\tau_i$ to denote a task's jobs and use the term \textit{task} and \textit{job} interchangeably.
  We also denote $d_i$ as the absolute deadline of $\tau_i$ (\ie deadline of any given job of $\tau_i$). 
%and let  $D_{min}$ be the smallest relative deadline of any task in $\Gamma$, \eg $D_{min} \overset{\operatorname{def}}{=} \underset{\tau_i \in \Gamma}{\min} ~\lbrace D_i \rbrace$. 
We assume cache related preemption delay is negligible compared to WCET of the tasks. We do not consider any precedence or synchronization constraints among tasks and $C_i, T_i, D_i \in \mathbb{N}^+,$.
 We further assume that the tasks have constrained-deadlines, \ie $D_i \leq T_i$ and the taskset is \textit{schedulable} by the EDF scheduling policy, that is the worst-case response time (WCRT)\footnote{The calculation of WCRT is presented in Section \ref{sec:radom_prio_inv}.} of each task is less than its deadline -- since \papername will be trivially ineffective for an unschedulable taskset.

% . Since the taskset is schedulable, the following (necessary and sufficient) condition will hold \cite{dbf_sanjay}:
% %\begin{equation} \label{eq:dbf}
% $\sum\limits_{\tau_i \in \Gamma} \mathsf{DBF}(\tau_i, t) \leq t,~\forall t \geq 0$
% %\end{equation}
% %In the above equation, 
% where the \textit{demand bound function} $\mathsf{DBF}(\tau_i, t)$ computes the cumulative maximum execution requirements of all jobs of $\tau_i$ (each of whose release time and deadline are within the interval $t$) and defined as follows:
% %\begin{equation}
% $\mathsf{DBF}(\tau_i, t) \overset{\operatorname{def}}{=} \max \left( 0, \left( \left \lfloor \frac{t- D_i}{T_i} \right\rfloor + 1 \right) C_i \right)$.
% %\end{equation}

Under the periodic task model, the schedule produced by any preemptive
scheduling policy, for a periodic taskset, is cyclic \ie the system will repeat the task arrival 
pattern after an interval that coincides with the taskset's \textit{hyperperiod}\footnote{The hyperperiod of the taskset is the least common multiple (LCM) of the periods of the tasks.} \cite{hyperperiod}, denoted by $L$.  
%The hyperperiod $L$ of the taskset $\Gamma$ is the least common multiple (LCM) of the periods of the tasks and defined as follows: $L \overset{\operatorname{def}}{=} \mathsf{lcm}(T_1, T_2, \cdots ,T_n)$.\footnote{For a set of positive real numbers $\mathcal{X} = \lbrace x_i \rbrace$, LCM is recursively defined as follows: $\mathsf{lcm}(\mathcal{X}) = \mathsf{lcm}(x_i, \mathsf{lcm}(\mathcal{X} \backslash x_i))$ where $\mathsf{lcm}(a, b) = \operatorname{inf} \lbrace y \in \mathbb{R}^+: \exists p, q \in \mathbb{N}^+ ~y = pa = qb \rbrace$.}
\begin{comment}
\hl{Furthermore, a discrete time model}~\cite{isovic2001handling} \hl{is considered.
We assume that a unit of time is integral. 
All system and task parameters are
multiples of a time unit. 
We denote an interval starting from time point $a$
and ending at time point $b$ that has a length of $b-a$ by $[a,b)$ or $[a,b-1]$.}
\end{comment}
Furthermore, we consider a  discrete time model (\eg integral time units~\cite{isovic2001handling}) where system and task parameters are
multiples of a time unit\footnote{We denote an interval starting from time point $a$
and ending at time point $b$ that has a length of $b-a$ by $[a,b)$ or $[a,b-1]$.}.

% \hl{Talk about cache related preemption delay in sys model/discussion}

% \hl{motivational example with a specific taskset  -- also clearly mention that this is a specific example and we did through evaluation}

% \hl{Talk about how entropy makes attack harder? Tie with the example CY generated -- the window is larger so attacker need to look a more time -- place in discussion?}

\subsection{Adversary Model}
\label{subsec::adversary}

We assume that the attackers have access to the timing parameters of the tasksets 
and also have knowledge of which scheduling policy is being used. The adversary's objective
is to get detailed information about the execution patterns of the real-time 
tasks
%, say via \textit{reconnaissance attacks} \cite{chen2015scheduleak}, 
and cause greater damage \cite{chen2015scheduleak,chen2018scheduleak}, to the system by exploiting the precise schedule information. %~\cite{chen2015scheduleak}. 
%Often times, the goal of the attacker goal may introduce catastrophe into the victim system.

As introduced in Section~\ref{sec::intro}, 
the attacker may exploit some side-channels (\eg power consumption, schedule preemptions, electromagnetic (EM) emanations and temperature) to observe and reconstruct the system schedule \cite{chen2015scheduleak}.
A smart attacker possessing sufficient system information can carry out more advanced  attacks under the right conditions, to move the system to an unsafe state.  
For example, in the now famous Stuxnet attack~\cite{stuxnet}, the malware was remnant in the system for \textit{months} to collect sensitive information before the main attack. 
%
%In this paper, we focus on the type of attacks that utilizes schedule information. 
%The attacker may exploit some side-channels to observe (and reconstruct) the system schedule. This may lead to attacks that \textit{exploit the determinism of the real-time schedule} to infer the exact instant a task starts to execute~\cite{chen2015scheduleak}.
%While the leakage of such schedule information may expose sensitive information, however the more catastrophic outcomes is that the information gained could also be used to disrupt critical tasks in the system. 
%%By knowing the precise timing information of the tasks, an attacker can launch further attacks at the right timing to increase the chance of success (often with high probability) of destabilizing the system.
It is possible for a denial-of-service attack to target only a specific service handled by a critical task when the precise schedule information is obtainable.
A side-channel attack~\cite{kelsey1998side_short, page2002theoretical_short} is also another typical class of attacks that can benefit from such schedule reconstruction
attacks.  
\begin{comment}
For example, a cache side-channel attack consists of two steps:
\textit{prime} and \textit{probe}~\cite{osvik2006cache_short}.  To launch such an attack against a task (\ie a victim task), the attacker first needs to fill the
cache memory right before the victim task starts.  Then, the attacker yields
and allows the victim task to run.  The attacker comes back to probe the cache
usage right after the victim task finishes its job.  In this process, the
closer the priming step is to the victim task's start time, the less noise in
the final attack results.  In other words, the success of the cache timing
attack greatly depends on the knowledge of the task's start time.
%\todo[inline]{Should we say something about covert channels?}
\end{comment}
For example, it was shown that the precise schedule information can be exploited to assist in determining the \textit{prime} and \textit{probe}~\cite{osvik2006cache_short} instants in a cache side-channel attack to increase the chance of success~\cite{chen2018scheduleak}.

We further assume that the scheduler is not compromised and the attacker does
not have access to the scheduler. Without this assumption, the attacker can
undermine the scheduler or directly obtain the schedule information. 
%This is a reasonable assumption since the scheduler operates with administrative privileges and attacks at the application level are easier. 
%Security mechanisms that prevent unauthorized access to the root privileges can improve the security of the scheduler.
Our objective, then, is to reduce the inferability of the schedule for real-time tasksets (and also reduce possibility of other attacks that depend of predictable schedules) while meeting real-time guarantees.
The randomness introduced to the schedule increases variations in the system and hence makes attacks that rely on the determinism of the real-time schedule, harder.
% \section{Bounding}
% Bound the probability of a task appearing in a slot

% Bound the number of context switches before and after applying EDF randomization.

\section{Schedule Randomization Protocol}
\label{sec:protocol_overview}

In this section we describe  the \papername protocol.
%whose objective is to reduce the inferability of the schedule for real-time tasksets. 
The focus of our design is such that, even if an observer is able to capture the exact schedule for a period of time (for instance, for a few hyperperiods), \papername will schedule tasks in a way that succeeding execution traces will show different orders (and timing) of execution for the tasks. The main idea is that at each scheduling point, we \textit{pick a random task from the ready queue} and schedule it for execution. However such random selection may lead to priority inversions \cite{priority_inversion_sanjay} and any arbitrary selection may result in \textit{missed deadlines} – hence, putting at risk the safety of the system. \papername solves this problem by allowing \textit{bounded priority inversions}. It restricts how the schedule may use priority inversions without violating real-time constraints (\eg deadline) of the tasks. To ensure this, \papername calculates an ``acceptable'' priority inversion budget. If the budget is exhausted during execution, then we stop allowing lower priority tasks to execute ahead of the higher priority task that has the empty budget. The following sections present the details of the \papername protocol.

\subsection{Randomization with Priority Inversion} \label{sec:radom_prio_inv}

A key step that is necessary for randomization is to calculate the maximum amount of time that lower priority jobs, $lp(\tau_i)$, can execute before $\tau_i$. 
This is much harder in EDF compared to the fixed-priority system (that prior work, TaskShuffler~\cite{taskshuffler}, was focused on) due to the dynamic nature of EDF (\ie the task priority varies at run-time). 
Therefore we define the \textit{worst-case inversion budget} (WCIB) $V_i$ that represents the maximum amount of time for which a job of some task $\tau_i$ with relative deadline $d_i$  may be blocked by a job of some task $\tau_j \in \Gamma, j \neq i$ with $d_j > d_i$ (and hence lower relative priority than $\tau_i$). In the following we illustrate how we calculate WCIB for each task by utilizing the response time analysis \cite{edf_wcrt_1, edf_wcrt_2} for EDF.

\subsubsection{Bounding Priority Inversions}

The WCRT of $\tau_i$ is the maximum time between the arrival of a job of $\tau_i$ and its completion. Our idea of bounding priority inversions is to calculate the slack times for each task (\eg difference between deadline and response time) and allow low priority tasks to execute up to that amount of time. We therefore define the WCIB of $\tau_i$ as follows:
% Based on the above calculations, we can derive the WCIB of $\tau_i$ as follows:
\begin{equation}\label{eq:wcib}
%$
V_i = D_i - R_i.
%$.
\end{equation}
where $R_i$ represents an upper bound of WCRT (see Appendix~\ref{sec:wcrt_calculation} for the calculation of $R_i$). \textit{The $V_i$ represents  the maximum amount of time for which all lower priority jobs $lp(\tau_i)$ (\eg $d_j > d_i$) are allowed to execute while an instance of $\tau_i$ is still unfinished without missing its deadline, even in the worst-case scenario.} 
The \papername protocol guarantees that the real-time constraints are satisfied by bounding priority inversions using $V_i, \forall \tau_i \in \Gamma$. 
Note that WCIB can be negative for some $\tau_i$ -- although non-positive WCIB does not attribute that the taskset is unschedulable. At each scheduling point $t$, our idea is to execute some low priority job $\tau_j$ with $V_j > 0$ up to $\min (\widehat{C}_j^t, V_j)$ additional time-units before it leaves the processor for highest priority job where $\widehat{C}_j^t$ represents the remaining execution time of $\tau_j$ at $t$.

% \hl{TODO: talk about $v_i$ when jobs finish earlier and also we are not constrained by the WCET of the tasks  -- MH}

We enforce the WCIB at run-time by maintaining a per-job counter, \textit{remaining inversion budget} (RIB) $v_i$, $0 \leq v_i \leq V_i$. RIB is initialized to $V_i$ upon each activation of the jobs of $\tau_i$ and decremented for each time unit when $\tau_i$ is blocked by any lower priority job. When $v_i$ reaches zero no job with absolute deadline greater than $d_i$ is allowed to run until $\tau_i$ completes. Note that not all the jobs of $\tau_i$ may need $C_i$ time unit for computation (recall that $C_i$ is the worst-case bound of the execution time). If some low-priority job $\tau_j$ (\eg $d_j > d_i$) that blocks $\tau_i$ finishes earlier than $C_j$, the RIB $v_i$ will not be decreased accordingly.

%Recall that the worst-case interference that $\tau_i$ can experience is bounded by $\max \lbrace I_i(a) \rbrace, ~0 \leq a < \widehat{R} - C_i$. Besides, 

For a given non-negative WCIB, jobs of $\tau_i$ can be delayed for up to $V_i$ by priority inversions. The WCRT of $\tau_i$ is bounded by $R_i + V_i = R_i + D_i - R_i = D_i$.
Hence, $\tau_i$ is schedulable with the \papername protocol and 
we can assert the following:

\begin{proposition}
If $\Gamma$ is schedulable under EDF,  
 WCIB is non-negative for some $\tau_i$ and low priority jobs of $\tau_i$ do not delay $\tau_i$ more than $V_i$ then \papername will not violate the real-time constraints of $\tau_i$.
\end{proposition}

\subsubsection{Selection of Candidate Jobs for Randomization}
As we mentioned earlier, when the run-time counter RIB (\ie $v_i$) reaches zero, no jobs with deadline greater than $d_i$ can run while $\tau_i$ has an outstanding job. However, lower priority jobs could cause $\tau_i$ to miss its deadline by inducing the worst-case interference from the higher priority jobs, \ie $\forall d_j < d_i$, due to the chain reaction. Therefore, to preserve the schedulability of such jobs we must prevent it from experiencing such additional delays. We achieve this by the following inversion policy:
%As we mentioned earlier, some task $\tau_i$ may have a negative WCIB, \ie $V_i < 0$. This will result the counter RIB $v_i$ always be negative and no lower priority jobs with deadline greater than $d_i$ can run while $\tau_i$ has an outstanding job. However, lower priority jobs could cause $\tau_i$ to miss its deadline by inducing the worst-case interference from the higher priority jobs, \ie $\forall d_j < d_i$ due to the chain reaction. Therefore, to preserve the schedulability of such jobs we must prevent it from experiencing such additional delays. We achieve this by defining the following inversion policy:

%{\noindent  $\blacktriangleright$  \bf \em Priority Inversion Policy (PIP): }{\em If WCIB $V_i < 0$ for some $\tau_i \in \Gamma$, no job $\tau_j$ with $d_j > d_i$ is allowed to run while any of high priority job $\tau_k$ with $d_k < d_i$ has an unfinished job. }
{\noindent  $\blacktriangleright$  \bf \em Randomization Priority Inversion Policy (RPIP): }
{\em If RIB $v_i < 0$ for some $\tau_i \in \Gamma$, no job $\tau_j$ with $d_j > d_i$ is allowed to run while any of high priority job $\tau_k$ with $d_k < d_i$ has an unfinished job. }

In order to enforce RPIP at run-time, at each scheduling decision point, we now define the variable minimum inversion deadline $m_i^t$ for jobs of $\tau_i$ as follows: 
%\\
%\hl{CY: $V_j$ should be $v_j$ in Equation 7.}\\
%\hl{No, see my slack message -- MH}\\
%\hl{CY: Disagree, using $V_j$ is inconsistent to the description of the 2nd point in B.Step1 as well as the description under Equation 9.}
% \hl{Check if $V_j$ below is needed. fixed -- MH}
%\begin{equation}
$m_i^t = \min \lbrace d_j | \tau_j \in \mathcal{R}_\mathcal{Q}^t,~ d_j > d_i \wedge v_j < 0 \rbrace.$ 
%\end{equation}
where $\mathcal{R}_\mathcal{Q}^t$ is the ready queue at scheduling point $t$. When there is no such task as $\tau_j$, $m_i^t$ is set to an arbitrarily large (\eg infinite) deadline. 
%arbitrary maximum deadline
The variable $m_i^t$ allows us to determine which jobs to \textit{exclude} from priority inversions. That is, no job that has a higher deadline than $m_i^t$ can be scheduled as long as $\tau_i$ has an unfinished job. Otherwise, the job with relative deadline $m_i^t$ (not the job $\tau_i$) could miss its deadline. %As an example, consider the following taskset:

\begin{example} \label{ex:simple_taskset}
%\hl{This example needs to be revised to consider RIB $v_i$ rather than $V_i$. MH: this is okay -- since at $t=0$, both $Vj$ and $vj$ same}
The taskset $\Gamma_{ex1}= \lbrace \tau_1, \tau_2, \tau_3, \tau_4 \rbrace$ contains the following parameters:

\begin{center}\footnotesize
\begin{tabular}{ |c|c|c|c| } 
 \hline
 Task & $C_i$ & $T_i = D_i$ & $V_i$ \\
 \hline
 $\tau_1$ & $4$ & $10$ & $1$ \\ 
 $\tau_2$ & $1$ & $20$ & $-2$ \\ 
 $\tau_3$ & $1$ & $5$ & $-2$ \\ 
 $\tau_4$ & $2$ & $12$ & $-1$  \\ 
 \hline
\end{tabular}
\end{center}
At $t=0$, $d_1 = 10$, $d_2 = 20$, $d_3 = 5$, $d_4 = 12$. For notational convenience, let us denote $m_i^0$ as $m_i$. Hence $m_1 = 12$, $m_2 = \infty$, $m_3 = 12$ and $m_4= 20$. Therefore at $t=0$ the job $\tau_2$ and $\tau_4$ are not allowed to participate in priority inversion (since $d_2, d_4 > m_i, i \in \lbrace 1, 3 \rbrace$ and $\tau_1$, $\tau_3$ have not completed.
\end{example}

It can be shown that at any scheduling point $t$ we can enforce RPIP by only examining the inversion deadline of highest priority (\eg shortest deadline) job, $m_{HP}^t$ \cite{taskshuffler}. Hence, at each scheduling decision, \papername excludes all ready jobs from the selection that have higher deadline than $m_{HP}^t$.

\subsection{Overview of The Randomization Protocol} \label{subsec:protocol}

% We now present the overview of the \papername protocol. 

The \papername protocol selects a new job using the following sequence of steps (refer to Algorithm \ref{alg:sched_rand}
for a formal description) at every scheduling decision point. 

% \todo[inline]{CY: Some items are commented out here. IEEE format to be fixed.}

\begin{itemize}
\item {\em Step 1 (Candidate Selection):} At each scheduling point $t$, the \papername protocol searches for possible candidate jobs (that can be used for priority inversion) in the ready queue. Let us denote $\mathcal{R}_\mathcal{Q}^t$ as the set of ready jobs, $\tau_{HP} \in \mathcal{R}_\mathcal{Q}^t$ is the highest priority (\ie shortest deadline) job in the ready queue and $\mathcal{C}_\mathcal{L}^t$ represents the set of candidate jobs at some scheduling point $t$.

\begin{itemize}
%\item We first add the highest priority job $\tau_{HP} \in \mathcal{R}_\mathcal{Q}^t$ to the candidate list \hl{since $\tau_{HP}$ has the earliest deadline and hence is always considered}. If its counter RIB is zero or smaller (\ie $v_{HP} \leq 0$), then \papername moves to Step 2 (since priority inversion is not possible due to its inversion budget being non-positive).
\item We first check the RIB of the highest priority job $\tau_{HP} \in \mathcal{R}_\mathcal{Q}^t$. If the RIB is zero, then $\tau_{HP}$ is added to the candidate and \papername moves to Step 2 since priority inversion is not possible due to its inversion budget being non-positive. 

\item When RIB is non-negative (\ie $v_{HP} > 0$), we iterate through the ready queue and add the job $\tau_i \in \mathcal{R}_\mathcal{Q}^t$ to the candidate list $\mathcal{C}_\mathcal{L}^t$ if its deadline is less than or equal to $m_{HP}^t$ (\ie the minimum inversion deadline of the highest-priority job at scheduling point $t$). 
%\hl{CY: If $V_j$ is used in Equation 7, then it's possible that a job with $V_j>0$ and $v_j==0$ is added to $\mathcal{C}_\mathcal{L}^t$.}
\end{itemize}

\item {\em Step 2 (Randomizing the Schedule):} This step selects a random job from the ready queue for execution. The selected job will run until the next scheduling decision point $t^\prime$. We randomly pick a job $\tau_R$ from  $\mathcal{C}_\mathcal{L}^t$ and set the next scheduling decision point as follows: %The sequence of operations in this step are described in the following.

\begin{itemize}
\item If $\tau_R$ is the highest priority job in the ready queue, the next decision point $t^\prime$ will be either when the job finishes or a new job of another task arrives.

\item Otherwise, the next decision\footnote{Section \ref{sec::fine_grained_mode} presents another approach to trigger scheduling decisions.} will be made at when $\tau_R$ completes or the inversion budget expires, that is, 
\begin{equation} \label{eq:t_prime}
t^\prime  = t +  \min (\widehat{C}_R^t, \widehat{v} ),
\end{equation}
unless a new job arrives before time $t^\prime$ where 
%\begin{equation}
$\widehat{v} = \min(v_j | \tau_j \in \mathcal{R}_\mathcal{Q}^t \wedge d_j < d_R)$
%\end{equation} 
and
$\widehat{C}_R^t$ represents the remaining execution time of $\tau_R$. Note that the variable $\widehat{v}$ is always positive since every job with a higher priority than the selected job has some remaining inversion budget. Otherwise, $\tau_R$ would not have been added to the candidate list in Step 1.
\end{itemize}

\end{itemize}

We now use a simple example to illustrate our randomization protocol.

\begin{figure}[!t]
\centering
\advance\leftskip-0.5em
\includegraphics[width=2.6in]{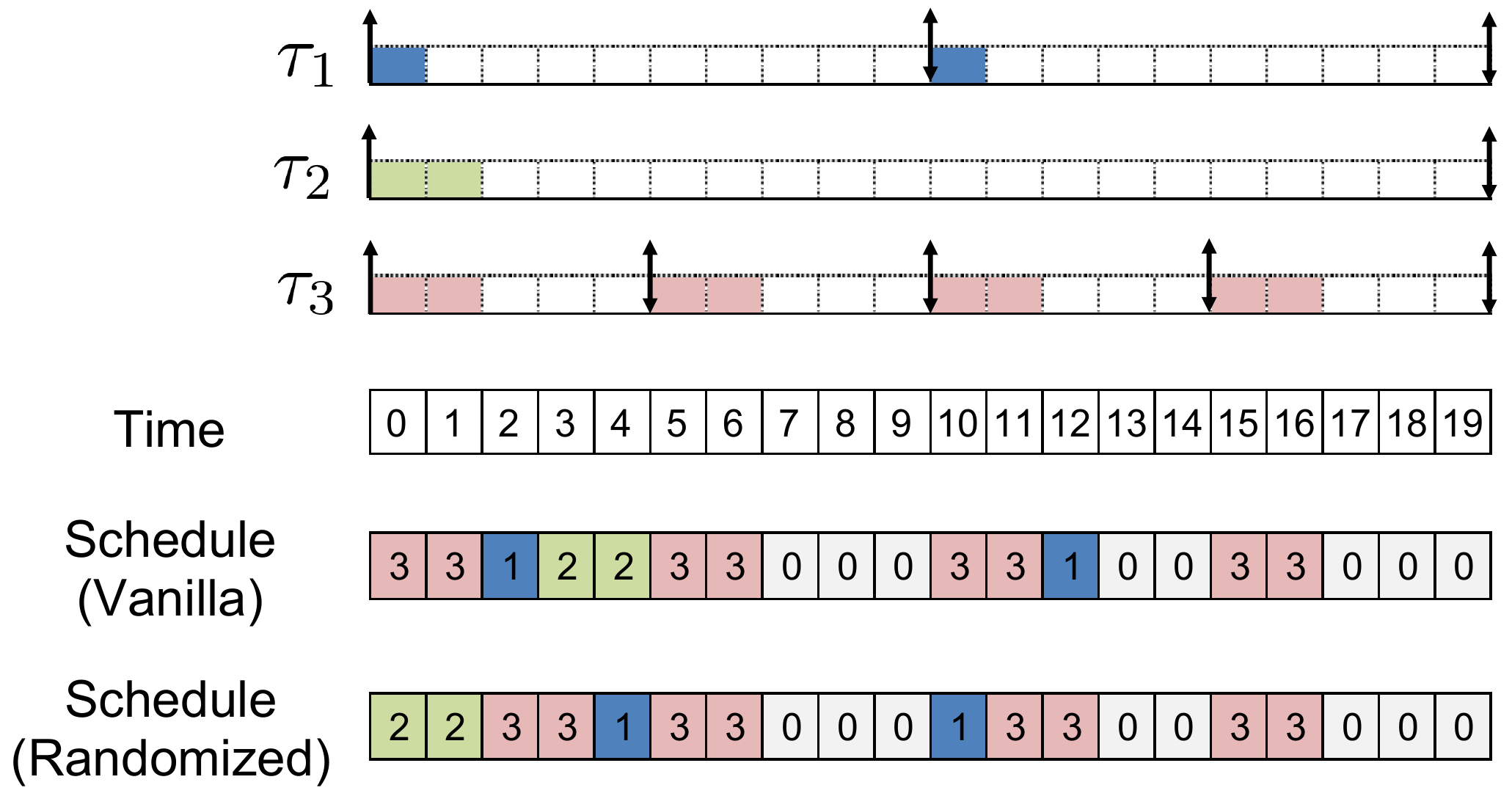}
\vspace{-0.4\baselineskip}
\caption{An instance of schedule randomization protocol for the taskset presented in Example~\ref{ex:simple_taskset2}. The length of the taskset hyperperiod is $L=20$ and $0$ represents idle time (\eg when no other tasks are active). The upward and downward arrows represent job activations and deadlines, respectively.} 
\label{fig:example}
\vspace{-1.5\baselineskip}
\end{figure}

\begin{example} \label{ex:simple_taskset2}
Let us consider the taskset $\Gamma_{ex2}= \lbrace \tau_1, \tau_2, \tau_3 \rbrace$ with following parameters:

\begin{center}\footnotesize
\begin{tabular}{ |c|c|c|c| } 
 \hline
 Task & $C_i$ & $T_i = D_i$ & $V_i$ \\
 \hline
 $\tau_1$ & $1$ & $10$ & $3$ \\ 
 $\tau_2$ & $2$ & $20$ & $5$\\ 
 $\tau_3$ & $2$ & $5$ & $3$ \\ 
 \hline
\end{tabular}
\end{center}

The taskset is schedulable by EDF (\eg $\sum\limits_{\tau_i \in \Gamma_{ex2}} \frac{C_i}{T_i} \leq 1$). The schedule of the vanilla EDF and an instance of randomization protocol is illustrated in Fig. \ref{fig:example}. %All the tasks are released at $t=0$. 
At time $t=0$, all three jobs are in ready queue and have positive inversion budget. Notice that $\tau_3$ is the highest priority job, $m_{HP} = \infty$ and all three jobs are in the candidate list. Let the scheduler randomly pick $\tau_2$ . From Eq. (\ref{eq:t_prime}), the next scheduling decision will be taken at $t = 0 + \min(2, 3) = 2$. At $t=2$, $v_1 = v_3 = 1$, $m_{HP} = m_3 = \infty$ and both $\tau_1$ and $\tau_3$ are in candidate list. Let $\tau_3$ be randomly scheduled (that is also the highest priority job). The next scheduling decision will be at $t = 2+2 =4$. At $t=4$, only $\tau_1$ is active and scheduled (next scheduling decision will be at $t=5$). $\tau_3$ is the only active job at time $t=5$ and hence scheduled. At $t=10$, both $\tau_1$ and $\tau_3$ are active, $m_3 = m_{HP} = \infty$ and hence both are in candidate list. $\tau_1$ is randomly scheduled and the next scheduling point will be at $t = 10 + 1 = 11$. At $t=11$, only $\tau_3$ is active and scheduled. Likewise, $\tau_3$ is the only active job at $t=15$ and scheduled. 

\end{example}

\subsection{Unused Time Reclamation}

As mentioned earlier, not all the jobs of a task may require worst-case unit of time for its computation. We propose to reclaim this unused time (\eg difference between WCET and actual execution time) to increase the inversion budget for lower priority jobs.
In the case that the (randomly) selected job finishes earlier (\ie the actual execution time is smaller than its WCET),  the unused time that is reserved for this job can be transferred to its lower priority jobs (\ie those ready jobs that have higher deadlines at the moment) as extra inversion budget.
Therefore, when enabling this feature, the RIBs of the lower priority jobs are updated (at the scheduling point $t^\prime$ when the selected job $\tau_R$ finishes its execution) as follows:
%\begin{equation}
$v_j = v_j + \delta_R^{t^\prime}, ~\tau_j \in \mathcal{R}_\mathcal{Q}^{t^\prime} \wedge d_j > d_R$
%\end{equation}
where $\delta_R^{t^\prime}$ represents the unused time over WCET and $\mathcal{R}_\mathcal{Q}^{t^\prime}$  is the ready queue at time $t^\prime$.
Note that the real-time constraints (\ie deadlines) are respected since Eq.~\ref{eq:wcib} for every ready job at time $t^\prime$ still holds (\ie $V_j + \delta_R^{t^\prime} = D_j - (R_j - \delta_R^{t^\prime})$) with the unused time transferring.%(\ie $v_j + \widehat{C}_R^{t^\prime} == D_j - (R_j-\widehat{C}_R^{t^\prime})$) \hl{you mean equal? or condition checking as in say C programming language?}

When there are no tasks in the ready queue (\eg during slack time), the processor is \textit{idle}, \eg nothing is executing in the system. Although \papername brings variations between the hyperperiods when compared to the vanilla EDF, randomizing only real-time tasks results in the schedule being somewhat predictable since the idle times (\ie slack) appear in nearly same slots. We address this problem by \textit{scrambling the idle times} along with the real-time tasks in the next section.

\subsection{Idle Time Scheduling}
\label{sec::idle_mode}
One of the limitations of randomizing only the tasks is that the task execution is squeezed between the idle time slots and the latter remain predictable. The work-conserving nature of EDF causes separations between task executions and idle times. Hence some tasks appear at similar places over multiple hyperperiods. One way to address this problem and improve schedule randomness is to \textit{idle the processor}, intentionally, at random times \cite{taskshuffler}. We achieve this by considering idle times as instances of an additional task, referred to as the \textit{idle task}, $\tau_I$. Then, the randomization protocol can be applied over the \textit{augmented taskset} $\Gamma^\prime = \Gamma \cup \lbrace \tau_I \rbrace$. 

%The schedule generated for $\Gamma_{ex}^\prime = \Gamma_{ex} \cup \lbrace \tau_I \rbrace$ is illustrated in Fig. \ref{fig:grid_withidle} where $0$ represents $\tau_I$. As the figure shows the tasks are spread over wider ranges compared to Fig. \ref{fig:grid_withoutidle}.

It can be noted that $\tau_I$ has infinite period, deadline and execution time, and hence always executes with the \textit{lowest} priority. 
Hence $\tau_I$ can force all other tasks $\tau_i \in \Gamma$ to maximally consume their inversion budgets. 
During randomization the idle task will convert a work-conserving schedule to a non-work-conserving one, but it will not cause any starvation for other tasks. This is because Step 2 of the \papername protocol (see Section  \ref{subsec:protocol}) selects candidate tasks in a way that real-time constraints for \textit{all} tasks in the system will always be respected. Randomizing the idle task effectively makes tasks appear across wider ranges and thus reduces predictability. As a result, the schedule can be less susceptible to attacks that depend on the predictability of RTS.

\subsection{Fine-Grained Switching}
\label{sec::fine_grained_mode}
In prior work \cite{taskshuffler} researchers proposed to decrease the inferability of the fixed-priority scheduler by randomly \textit{yielding} a job, early, during execution. As a result the schedule will be fragmented at different time-points and thus will bring more variations across execution windows. Our proposed \papername protocol can also be modified to incorporate such a feature. Recall that the scheduling decisions in our scheme are made either when: \ci a new job arrives, \cii a job completes, or \ciii the inversion budget expires (refer to Step 2 in Section \ref{subsec:protocol}). Therefore we can achieve fine-grained switching by modifying the next scheduling decision point $t^\prime$ in Eq. (\ref{eq:t_prime}) as follows:
%\begin{equation} \label{eq:t_prime_fine}
$
t^\prime = t + \operatorname{rand}(1,  \min (\widehat{C}_R^t, \widehat{v} ))
$
%\end{equation}
where the function $\operatorname{rand}(a, b)$ outputs a random number between $[a,b]$. 

% In Fig. \ref{fig:grid_fine} we illustrate the schedule of $\Gamma_{ex}^\prime$ using the fined-grained randomization scheme. While some tasks are completed earlier (for this specific taskset and randomization instance) and leaving the idle tasks for later slots, it does not attribute that this scheme is inferior than the previous one. 

\subsection{Algorithm}
\label{sec:algo_develop}

Algorithm \ref{alg:sched_rand} formally presents the proposed schedule randomization protocol. This event-driven algorithm executes at the scheduler-level and takes the taskset (with idle time) $\Gamma^\prime = \Gamma \cup \lbrace \tau_I \rbrace$ as an input. %operates in an iterative manner. 
At each scheduling decision point $t$, a ready job is (randomly) selected for scheduling and the next scheduling decision point $t^\prime$ is determined.

%\todo[inline]{CY: For line 22 in Algorithm 1, let's match Equation 8. Meaning line 22 to be $\Delta t := \min (\widehat{C}_R^t, \widehat{v} )$. Note that my program implements Equation 8. 
%MH: please have a look in my Slack message.}

%\todo[inline]{Pending - Wait for CY's verification \\
%CY: Line 3 in the following algorithm, the condition $\exists \tau_j \in \mathcal{R_Q} :: d_j - t < D_{min}$ will make the the task of $D_{min}$ be picked all the time as long as it is in the ready queue (except at its arrival instant). This condition should simply be $\exists \tau_j \in  \mathcal{R_Q} :: \mathsf{IBF}(d_j -t)=0$.}
 \renewcommand\algorithmiccomment[1]{%
 {\it /* {#1} */} %
}
\renewcommand{\algorithmicrequire}{\textbf{Input:}}
    \renewcommand{\algorithmicensure}{\textbf{Output:}}
    \renewcommand{\algorithmicforall}{\textbf{for each}}
    
		\begin{algorithm}[!t]
        %[H]
%         \footnotesize
			\begin{algorithmic}[1]
				\begin{footnotesize}
				\REQUIRE Augmented task set $\Gamma^\prime = \Gamma \cup \lbrace \tau_I \rbrace$ and current scheduling point $t$
    \ENSURE The randomized schedule $S_t$ and the next scheduling point $t^\prime$
					\vspace{0.4em}
                    
                    %\STATE $t \leftarrow$ current scheduling point
                    
                \STATE $\mathcal{R}_\mathcal{Q}^t :=$ set of ready jobs     
                    
           \STATE Add the highest priority job to the candidate list, \ie $\mathcal{C}_\mathcal{L}^t := \lbrace \mathcal{R}_\mathcal{Q}^{HP} \rbrace$
           \STATE \COMMENT{Search candidate jobs if the highest priority job has non-zero inversion budget}
           \IF{$v_{HP} > 0$}
             \FORALL{$\tau_j \in \mathcal{R}_\mathcal{Q}^t$}
               \IF{$d_j \leq m_{HP}^t$}
                  \STATE $\mathcal{R}_\mathcal{Q}^t := \mathcal{R}_\mathcal{Q}^t \cup \lbrace \tau_j \rbrace$ \COMMENT{add $\tau_j$ to candidate list}
               \ENDIF
             \ENDFOR
           \ENDIF

           \IF{ $\mathcal{C}_\mathcal{L}^t = \lbrace \mathcal{R}_\mathcal{Q}^{HP} \rbrace$ }
           \STATE \COMMENT{schedule the highest priority (shortest deadline) job}
           \STATE $S_t := \mathcal{R}_\mathcal{Q}^{HP}$
           \STATE Set next scheduling point $t^\prime :=$ when new job arrives or current job completes
           \ELSE
           \STATE \COMMENT{randomly select a job $\tau_R$ from $\mathcal{C}_\mathcal{L}^t$} \STATE $S_t := \tau_R$
           
           \IF{$\tau_R = \mathcal{R}_\mathcal{Q}^{HP}$}
           \STATE Set next scheduling point $t^\prime :=$ when new job arrives or current job completes
           \ELSE 
           \STATE \COMMENT{set the next random scheduling point $t^\prime$ as a function of current job completion or budget expiration time (unless a new job arrives before $t^\prime$)}
           \STATE $\Delta t := \operatorname{rand}(1,  \min (\widehat{C}_R^t, \widehat{v} ))$
           \STATE Set next scheduling point $t^\prime  := t +  \Delta t$
           \ENDIF
           \ENDIF
           \STATE \COMMENT{return the scheduled job and the next scheduling point}
           \RETURN $(S_t, t^\prime)$
                   	\end{footnotesize}
				\end{algorithmic}
                
                \caption{Schedule Randomization Protocol}
			\label{alg:sched_rand}
                \end{algorithm}

% At each scheduling point $t$, 
In Lines 3-10, the algorithm first selects the set of candidate jobs $\mathcal{C}_\mathcal{L}^t$ using the procedure described in Section \ref{subsec:protocol} (see Step 1). If the highest priority job $\mathcal{R}_\mathcal{Q}^{HP}$ has negative inversion budget (\eg $v_{HP} \leq 0$), it will be scheduled for execution (Line 13). Otherwise it schedules a random job from the candidate list (Line 17). If the selected job is the highest priority job, the next scheduling point $t^\prime$ is set when the job completes or a new job of another task arrives (Line 14 and 19). If the selected job is not the highest priority one, the algorithm selects $t^\prime$ when the current inversion budget expires, unless the job completes or a new job arrives before $t^\prime$
%using Eq. (\ref{eq:t_prime_fine}) 
(Line 23).

The algorithm iterates over the jobs in the current ready queue $\mathcal{R}_\mathcal{Q}^t$ once and makes a single draw from the candidate list $\mathcal{C}_\mathcal{L}^t \subseteq \mathcal{R}_\mathcal{Q}^t$. 
%\hl{CY: We should also mention that computing the minimum inversion deadline is also another iteration of $|\mathcal{R}_\mathcal{Q}^t|$ and consider to put this step in Algorithm 1.}
Assuming a single draw from a uniform distribution (Line 17 and 22) takes no more than $O(|\mathcal{R}_\mathcal{Q}^t|)$, the complexity\footnote{Section~\ref{sec:linux_result} presents empirical evaluations for scheduling overhead.} of each instance of the algorithm is 
$O(|\mathcal{R}_\mathcal{Q}^t|)$ %where $|\cdot|$ denotes the cardinality of set $(\cdot)$.
.
\section{Schedule Entropy: A Measure of Randomness}
\label{sec:entropy}

While the mechanisms presented in Algorithm \ref{alg:sched_rand} obfuscates the inherent determinism in conventional dynamic-priority schedules, we still need to \textit{quantify the randomness} that has been introduced into the schedule. This can be addressed by 
analyzing the \textit{schedule entropy} that measures the randomness (or unpredictability) in the real-time schedule. Since prior entropy calculations do not capture the randomness of a schedule correctly (refer to Appendix \ref{subsec:entropy_limitation} for details) we now introduce a better approach to measure the schedule entropy.

%In the following, we first describe the limitations of existing entropy calculation techniques \cite{taskshuffler} and then introduce a better approach to measure the schedule entropy, followed by an evaluation of the entropy of the various schemes using synthetic workloads.

\subsection{Entropy of a \papername Schedule}
\label{subsec::reorder_entropy}

The proposed concept is based on a statistical model -- \textit{approximate entropy (ApEn)} \cite{apen} that is used to evaluate the amount of regularity in time series data. Let us consider $K > 1$ hyperperiods for a taskset $\Gamma$ (with hyperperiod-length $L$) that is represented as $K$ vectors of length $L$ as follows: $ [s_0^{1},\cdots,s_{L-1}^{1}],\cdots, [s_0^{K},\cdots,s_{L-1}^{K}]$. Each vector includes $L$ intervals of length $m$ of the form $X=\left[s_{t \!\!\! \mod L},s_{(t+1)\!\!\!\mod L},\cdots,s_{(t+m-1)\!\!\!\mod L}\right]$, $0\le t\le L-1$ and hence, we have total $\lambda=KL$ number of intervals of length $m$. Let us consider
%\begin{equation}
$X_t^{k}(m)=\left[ s^{k}_{t \!\!\!\mod L},s^{k}_{(t+1) \!\!\! \mod L},\cdots ,s^{k}_{(t+m-1) \!\!\! \mod  L} \right]$
%\end{equation}
as the interval of size $m$ starting from $s_t^{k}$ on the $k$-th hyperperiod where $ 0 \le t \le L-1$ and $1 \le k \le K$. %For given values for window size $1 \le m \le L$, and threshold $r$, our randomness measure is therefore calculated as follows:
For all intervals $X^{(k)}_t(m)$, let us define the following variable:
%\begin{equation}
$
C_t^{k}\coloneqq\frac{1}{K}\left|\{k': \delta(X^{k}_t(m),X^{k^\prime}_t(m))\le \pi,~1\le k^\prime \le K\}\right|,
$
%\end{equation}
where $\delta(\cdot,\cdot)$ denotes the \textit{dissimilarity}  between two intervals of different hyperperiod, $\pi$ is a given dissimilarity threshold and $|\cdot|$ represents the set cardinality.
We use Hamming distance \cite{hamming_distance} to evaluate the dissimilarity between intervals -- since this a relatively simple and widely used dissimilarity measure. For two vectors $U = [u_i]_{1 \leq i \leq m}$ and $V = [v_i]_{1 \leq i \leq m}$ of size $m$, Hamming distance is calculated as follows:
%\begin{equation}
$\delta(U, V)=\sum\limits_{i=1}^m \mathbb{I}(u_i \neq v_i)$,
%\end{equation}
where $\mathbb{I}(\cdot)$ is the indicator function that equals $1$ if the condition $(\cdot)$ is satisfied or $0$ otherwise. Notice that, $C_t^{k}$ represents the number of  intervals of length $m$ starting from $s_t^{k^\prime}$, $1\le k^\prime \le K$ with dissimilarity (in terms of Hamming distance) less than or equal to $r$ from $X^{k}_t(m)$ and normalized by the number of observed hyperperiods (\ie $K$). 

Let us now define the variable $\eta_t$ as an estimation of the entropy of variable $X_t^k(m)$, \ie an estimation of the entropy of a vector that starts from slot $t$ with length $m$ as follows:
%\begin{equation} \label{eq:eta_t}
$\eta_t=-\frac{1}{K}\sum\limits_{k=1}^{K}\log_2 C^{k}_t.$
%\end{equation}
Therefore for a given interval length $1 \le m \le L$ and dissimilarity threshold $\pi$, the randomness  (entropy) of a schedule observed over $K$ hyperperiods is given by the following equation:
%\begin{equation}
$
\widehat{H}(\mathcal{S}^k, m,\pi,K)=\frac{1}{m}\sum\limits_{t=0}^{L-1}\eta_t.
$
Notice that, for a deterministic scheduler (\eg when \textit{all} the jobs of the tasks take WCET for computation for vanilla EDF) the schedule entropy $\widehat{H}(\mathcal{S}^k,m,\pi,K)$ will be equal to \textit{zero} (\ie there is no randomness, as expected). 

%\end{equation}
%where $\eta_t$ is given by Eq. (\ref{eq:eta_t}).  

%$\eta_t=-\frac{1}{K}\sum\limits_{k=1}^{K}\log_2 C^{k}_l.$

%Note that for $\pi=0$, choosing $m=1$ gives us $\widehat{H}(\mathcal{S}^k, m,\pi,K)=\widetilde{H}(\mathcal{S}^k)$ and choosing $m=L$ outputs $\widehat{H}(\mathcal{S}^k, m,\pi,K)\rightarrow{H}(\mathcal{S}^k)$, as $K\rightarrow\infty$, where $\widetilde{H}(\mathcal{S}^k)$ and $H(\mathcal{S}^k)$ are defined in Eqs. \eqref{eq:sumentropy} and \eqref{eq:entropy}, respectively. 
% Notice that, for a deterministic scheduler (\eg when \textit{all} the jobs of the tasks take WCET for computation for vanilla EDF) the schedule entropy $\widehat{H}(\mathcal{S}^k,m,\pi,K)$ will be equal to \textit{zero} (\ie there is no randomness, as expected). 

%\hl{Sibin: Move the results from synthetic tasks (entropy \etc) here (and also the description of taskset generation.)}

\subsection{Interpretation of Entropy}
\label{sec:entropy_intuition}
%EDF 6.12
%UTR 9.49
The schedule entropy depicts the randomness for a given schedule. When comparing the entropies of two schedule sequences with equal lengths, a higher value implies that more variations are introduced in each time slot and the chance for a task appearing at the same time slot in every hyperperiod is smaller.
Consider the taskset presented in Example \ref{ex:simple_taskset} as an example -- the schedule entropies are $6.12$ and $9.49$ when scheduled by vanilla EDF and \papername, respectively.
As we can see from the frequency spectrum of the two schedule sequences (Fig.~\ref{fig:fft}), the higher randomness reduces the determinism in the schedule and some periods become unidentifiable from the spectral analysis.

Other methods, such as side-channel attacks, also suffer since the victim
tasks can potentially appear in larger ranges of executions. Such attacks typically
require ``prepping'' (\eg prime and probe \cite{osvik2006cache_short}) of the system and the closer
this is done to the actual execution of the victim task, the better it is for the
adversary. With increasing entropy values, the attacker has lesser precision in narrowing the exact arrival times for the victim task(s) and hence, experiences more noise in
measurements. Similarly, covert channels \cite{leak2} will also suffer since the expected
execution order of tasks is broken due to the randomization -- hence, higher entropy
values result in larger variations from the ``expected'' covert channel.

%As a result, attacks such as \ci scheduler side-channel attacks exploiting the periodicity of a given taskset, \cii covert channels relying on the schedule's determinism, \ciii cache timing side-channel attacks requiring the task's regularity, become difficult to achieve their intended goals.
%the window time in which a task can appear 

%================subsection===================
\subsection{Evaluation of Schedule Entropy}
\label{sec:exp_syntehtic}
We now evaluate the \papername protocol with synthetic workloads. %\footnote{This is a standard approach followed by real-time community for broader design-space exploration.}.
This is to understand the degree of randomness introduced into the schedule and we use the schedule entropy calculations from Section \ref{subsec::reorder_entropy}.
%synthetic workloads for broader design-space exploration.
The evaluation of scheduling overhead on a real platform is presented in Section~\ref{subsec:linux}.

%Understanding the entropy

\subsubsection{Simulation Setup}\label{subsec:sim_setup}
We used the parameters similar to that in earlier research \cite{sg1,taskshuffler, linux_push_pull, sanjay_limited_edf}. The tasksets were grouped into base-utilization buckets (\eg total sum of the task utilizations) from $[0.01+0.1 \cdot i, 0.1+0.1 \cdot i]$ where $i \in \mathbb{Z} \wedge 0 \leq i < 9$. Each base-utilization group contained $250$ tasksets and each of which had $[3, 10]$ tasks. We only considered tasksets that were schedulable by EDF.% \ie tasksets for which the condition in Eq. (\ref{eq:dbf}) was satisfied.

For a given base-utilization bucket, the utilization $U_i$ of individual tasks were generated from a uniform distribution using the UUniFast \cite{uunifast} algorithm. The period of each task was greater than $10$ with a divisor of $100$. This allowed us to set a common hyperperiod (\eg $L = 100$) for all the tasksets. We assumed that the deadlines are implicit, \eg $D_i = T_i,~\forall \tau_i$. The execution time $C_i$ for each of the tasks in the taskset was computed using the generated period and utilization: $C_i = \lceil U_i  T_i  \rceil$. The execution time of each job of $\tau_i$ was randomly selected from $\lceil \alpha \cdot C_i \rceil$ where $\alpha = [0.5, 1]$. The interval window size was $m=\lceil 0.35 L \rceil$ and the dissimilarity threshold $\pi$, $0.1 L$ (Appendix \ref{appsec:tru_apen_cor}). For each schedulable taskset, we observed the schedule for $K=100$ hyperperiods.

% \begin{table}[!htb]
% \vspace{-0.1in}
% \renewcommand{\arraystretch}{1.2}
% \caption{Experimental Parameters}
% \label{tab:ex_param}
% \centering
% %\begin{tabular}{c||p{5.9cm}}
% \begin{small}
% \begin{tabular}{p{4.20cm}||p{3.30cm}}
% \hline %\\
% \bfseries Parameter & \bfseries Values\\
% \hline\hline
% Number of Tasks & $[3, 10]$ \\
% Period of the Tasks & $[10, 100]$ \\
% Interval Length & $35$ \\
% Dissimilarity Threshold & $10$
% \\
% \hline
% \end{tabular}
% \end{small}
% \vspace{-0.1in}
% \end{table}

\subsubsection{Results}

\begin{figure}[!t]
\centering
%\advance\leftskip-0.8em
\includegraphics[width=0.75\linewidth,keepaspectratio]{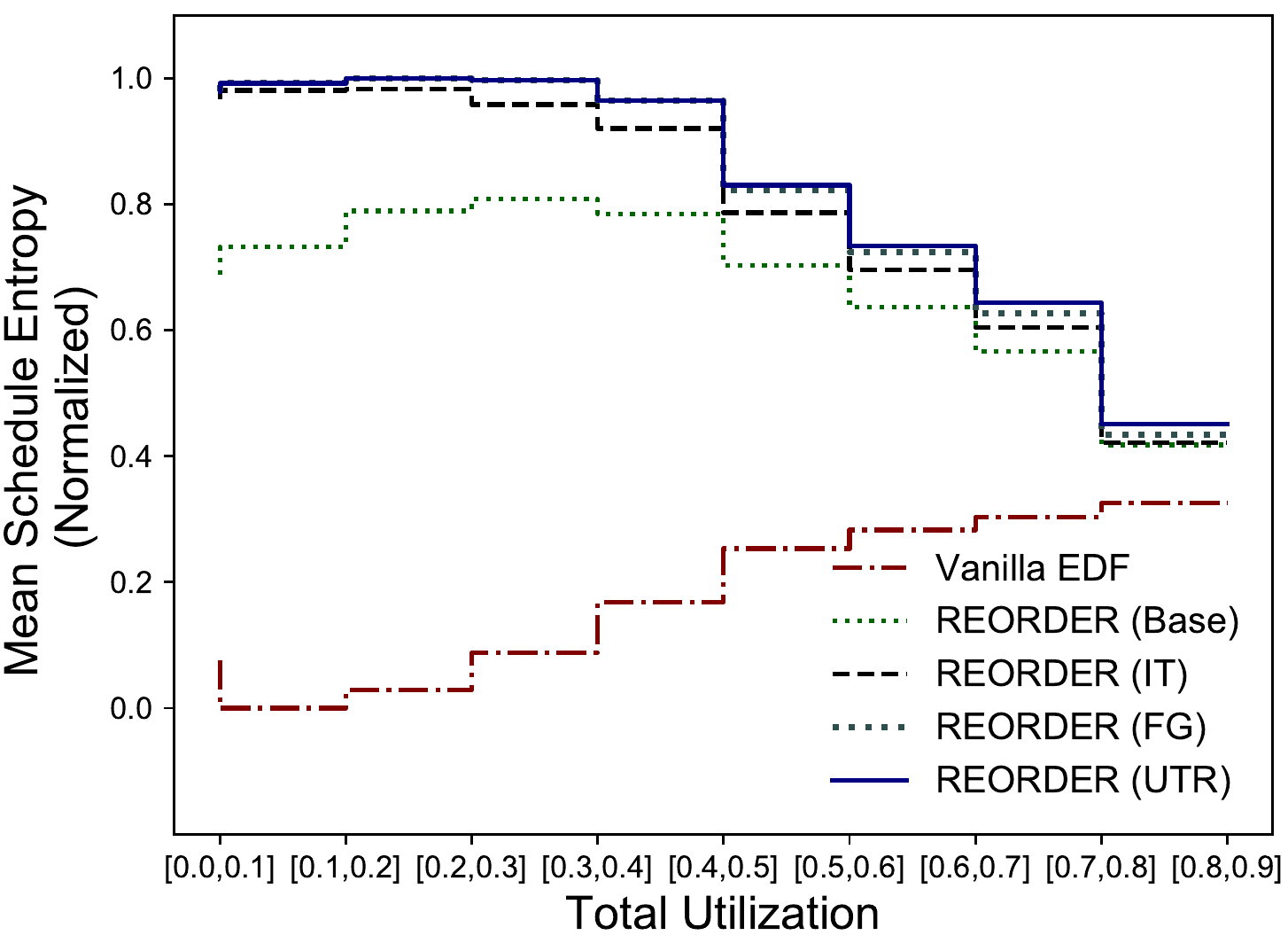}
\vspace{-0.5\baselineskip}
\caption{The average entropy of the system with varying total utilization for different randomization schemes. The \papername schedule shows maximum randomness (\eg entropy) in the low-to-medium base-utilization (\eg $< 0.7$).} 
\label{fig:entropy}
\vspace{-1.5\baselineskip}
\end{figure}

We now evaluate how much randomness (\viz~ unpredictability) the \papername protocol incurs {\em w.r.t.} vanilla EDF using the following schemes: 
\begin{itemize}
\item \papername (Base): only tasks are randomized; 
\item \papername (IT): randomization with augmented tasksets (\eg including idle time randomization); 
\item \papername (FT): fine-grained switching for augmented taskset (\eg yielding tasks at random points); and 
\item \papername (UTR): randomization with fine-grain scheduling and unused time reclamation.
\end{itemize}
% \ci \papername (Base): base randomization -- only tasks are randomized, \cii \papername (IT): randomization with augmented tasksets (\eg including idle time randomization), \ciii \papername (FT): fine-grained switching for augmented taskset (\eg yielding tasks at random points) and \civ \papername (UTR): randomization with fine-grain scheduling and unused time reclamation.

In these experiments we focus on observing the \textit{average behavior} of randomization schemes. In Fig. \ref{fig:entropy} we present the average schedule entropy of vanilla EDF (\eg no randomization) along with different randomization schemes. 
% \ci \papername (Base): base randomization -- only tasks are randomized, \cii \papername (IT): randomization with augmented tasksets (\eg including idle time randomization), \ciii \papername (FT): fine-grained switching for augmented taskset (\eg yielding tasks at random points) and \papername (UTR): \civ randomization with fine-grain scheduling and unused time reclamation.
% \hl{TODO: update descriptions as in new experiments.}

The X-axis of Fig. \ref{fig:entropy} shows the total system utilization. The Y-axis represents mean schedule entropy (normalized to $1$), \eg $\widehat{H}_{mean}(\cdot) = \tfrac{1}{\hat{n}_s} \sum\limits_{i=1}^{\hat{n}_s} \widehat{H}_i(\cdot)$, where $\hat{n}_s$ represents the  number of schedulable tasksets for a given base-utilization group and $\widehat{H}_i(\cdot)$ is the entropy of taskset $i$. 
%Recall that for vanilla EDF entropy is zero since the same schedule repeats in every hyperperiod.  
For higher utilizations entropy for vanilla EDF increases since the schedule across multiple hyperperiods becomes different because of less slack (\eg idle times). As we can see from this figure, the randomization protocol significantly increases schedule entropy. The idle time randomization with fine-grained scheduling and unused time reclamation (\eg \papername (UTR)) significantly improves the entropy over base randomization. Note that for higher utilization the improvement is marginal. This is due the fact that for higher utilization, the system does not have enough slack (\eg idle times) to randomize much -- and hence all three schemes show similar results (in terms of schedule entropy). 
%Another interesting observation from Fig. \ref{fig:entropy} is that the randomization protocol is most effective in the medium utilization\footnote{Similar trend  is also observed in prior work \cite{taskshuffler} for randomizing fixed-priority scheduling.} (\eg when the base-utilization is $[0.3, 0.7]$). This is because, for low utilization (\eg when the system has high slack) the processor is idle most of the times and the amount of time spent in task executions is relatively small. As a result the lesser number of variations (within the given interval length $m$) across hyperperiods reduces the average entropy. 
%As the load increases (\eg lesser amount slack), there are a larger number of candidates for randomization that results more variation across hyperperiods (\eg outputs higher entropy). However, 
As the utilization increases (\eg lesser slack), there are very few candidate jobs for priority inversions because of higher load. Hence, the entropy (\ie randomness) drops -- albeit the schedule is still less predictable compared to the vanilla EDF (since the mean entropy is greater than entropy of EDF).

\begin{figure}[!t]
\centering
%\advance\leftskip-0.8em
\includegraphics[width=0.85\linewidth,keepaspectratio]{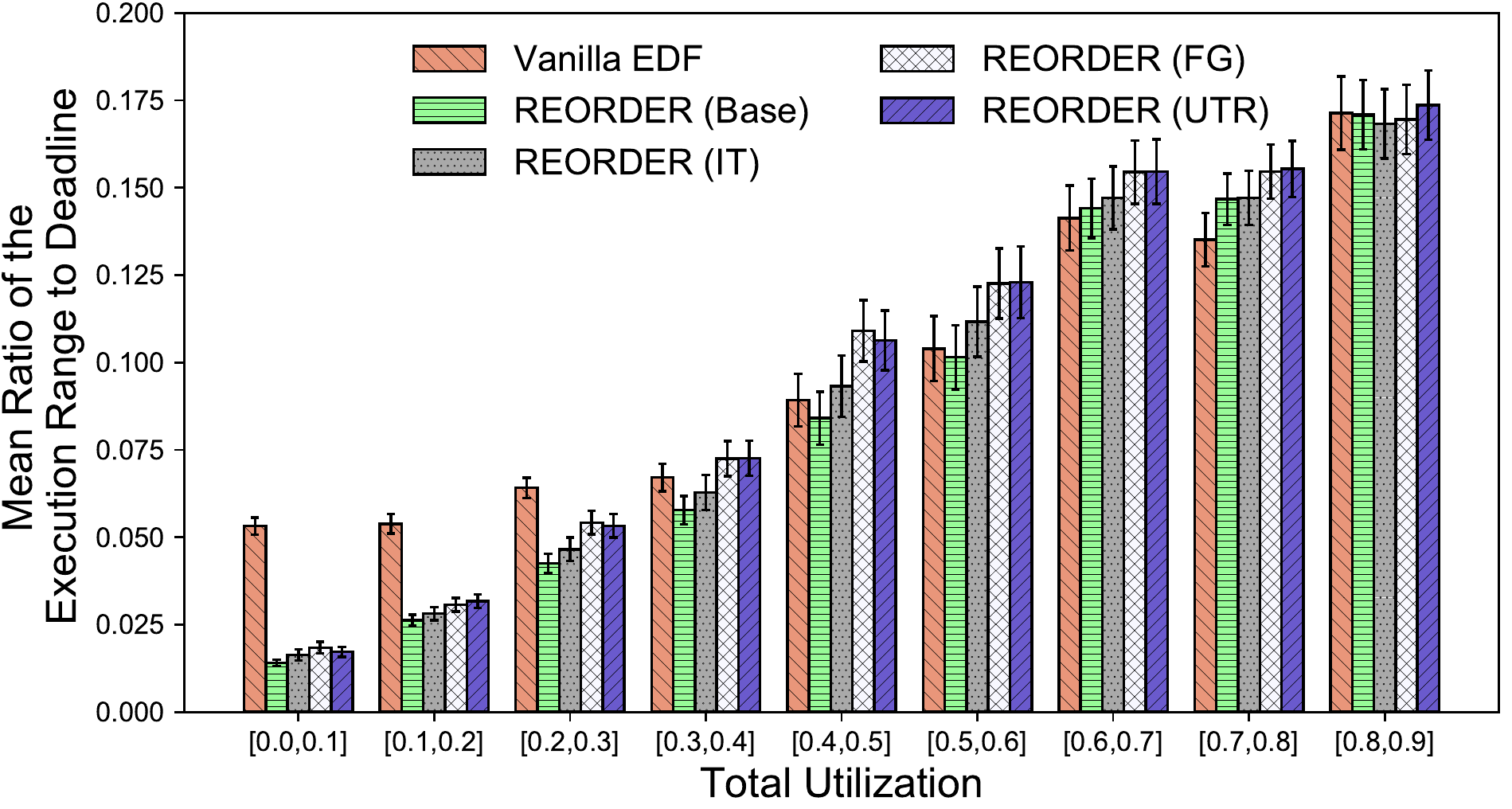}
\vspace{-0.5\baselineskip}
\caption{The geometric mean of the execution range to the deadline ratio. For \papername the mean ratio is higher when utilization is greater than $0.4$ -- that implies the tasks appear in wider ranges and hence it is harder to infer the actual execution time.} 
\label{fig:wcrt-ratio}
\vspace{-0.5\baselineskip}
\end{figure}

Another way to observe the schedule randomness is to measure the ranges within which each task can appear. A wider range implies  that is is harder to predict when a task executes. In this experiment we measured the first and the last time slots where a job of each task $\tau_i$ appears and used the difference between them as the range of execution for $\tau_i$ (denoted as $w_i$). In Fig. \ref{fig:wcrt-ratio} we show the ratio of execution range to deadline (\eg $\tfrac{w_i}{D_i} \leq 1$) of the tasks. The X-axis of the figure shows total utilization and Y-axis represents the geometric mean of the task execution range to deadline ratios in each taskset. 

%\hl{TODO UPDATE TEXT}
For low utilization situations, tasks appear within narrow ranges because of the work-conserving nature of the EDF algorithm. With increasing utilization the ranges become wider. This is because the worst-case response times of tasks (particularly, for lower priority tasks) increases due to the higher loads. For lower utilization, the system is dominated by slack times and hence randomizing tasks do not improve the execution range compared to vanilla EDF. This is because some (low-priority) jobs finish earlier due to priority inversions and hence the response time of those jobs is actually lower than the EDF scheme. As a result mean ratio for \papername decreases. As the figure shows, for higher utilization (\eg utilization greater than $0.4$) tasks appear in wider ranges (\eg higher mean ratio) when \papername is enabled. This is due to the fact that priority inversions with \papername (FT/UTR) increase task response times (especially for higher priority tasks). Besides, inverting the priority can also move lower priority jobs closer to their release times, thus widening the execution range. Since the tasks with \papername appear in wider ranges this also prevents an attacker from triggering side channel attacks as we mentioned in Section~\ref{sec:entropy_intuition}.

\section{Implementation}
\label{sec:implementation}
We implemented \papername in a real-time Linux kernel running on a realistic embedded platform to validate its usability and to evaluate its overhead.
To this end we also measure the overheads by comparing this to an existing vanilla EDF scheduler.
In this section we provide platform information and a high level overview of the implementation.
We have open-sourced our implementation and make it available on an anonymized public repository \cite{github:redf}. 
%We also integrated our randomization protocol into the real-time Linux kernel. 
The platform information and configurations are summarized in Table~\ref{tab:implementation_platform_summary}.
%The source code of the implementation is available on a public git repository at the link listed in the reference~\cite{github:redf}.

\begin{table}[t]%\small
\centering
\caption{Summary of the Implementation Platform}
\vspace{-0.5\baselineskip}
\label{tab:implementation_platform_summary}
\begin{tabular}{l||l}
\hline
% \multicolumn{2}{|c|}{\textbf{Platform Information}} 	\\ \hline
{\bfseries Artifact}         & {\bfseries Parameters} \\ \hline \hline
Platform         & ARM Cortex-A53 (Raspberry Pi 3) \\
System Configuration         & 1.2 GHz 64-bit processor, 1 GB RAM  \\ %\hline
Operating System & Debian Linux (Raspbian)							\\ %\hline
Kernel Version   & Linux Kernel 4.9.48 				\\ %\hline
Real-time Patch    & PREEMPT\_RT 4.9.47-rt37		\\ \hline %\hline
% \multicolumn{2}{|c|}{\textbf{Configurations}} 		\\ \hline
\begin{tabular}[c]{@{}l@{}}Kernel Configuration \\ ($\mathtt{make~defconfig}$)\end{tabular}   & \begin{tabular}[c]{@{}l@{}}$\mathtt{CONFIG\_SMP}$ disabled\\ $\mathtt{CONFIG\_PREEMPT\_RT\_FULL}$ enabled\end{tabular} \\ %\hline
Boot Commands & $\mathtt{maxcpus}$=1				\\ %\hline
Run-time Variables &  $\mathtt{sched\_rt\_runtime\_us}$=$-$1\\ 
 & $\mathtt{scaling\_governor}$=$\mathtt{performance}$ \\ \hline
 MiBench Applications & Security: $\mathtt{sha,blowfish}$ \\ 
    & Consumer: $\mathtt{typeset} $ \\
& Automotive: $\mathtt{basicmath,bitcount,}$ \\ 
   & \hspace{1.5cm}$\mathtt{qsort, susan}$ \\ \hline

\end{tabular}
\vspace{-2\baselineskip}
\end{table}

\subsection{Platform and Operating System}

We used a Raspberry Pi 3 (RPi3) Model B\footnote{\url{https://www.raspberrypi.org/products/raspberry-pi-3-model-b/}.}
development board as the base platform for our implementation.
The RPi3 is equipped with a 1.2 GHz 64-bit quad-core ARM Cortex-A53 CPU developed on top of  Broadcom BCM2837 SoC (System-on-Chip). RPi3 runs on a vendor-supported open-source operating system, \textit{Raspbian} (a variant of Debian Linux). We forked the Raspbian kernel and modified it (refer to the following sections) to implement the \papername protocol. Since we focus on the single core EDF scheduler in this paper, the multi-core functionality of RPi3 was deactivated by disabling the $\mathtt{CONFIG\_SMP}$ flag during the Linux kernel compilation phase. The boot command file was also set with $\mathtt{maxcpus}=1$ to further ensure the single core usage.

%The developers maintain the Raspbiran source code on a publicly available git repository. We forked the up-to-date repository to start our implementation. As a result, our implementation is based on the Linux kernel version 4.9.48.

\subsection{Real-time Environment}

%It is known that the 

The mainline Linux kernel does not provide any hard real-time guarantees even with the custom scheduling policies (\eg $\mathtt{SCHED\_FIFO}$, $\mathtt{SCHED\_RR}$). However
the \textit{Real-Time Linux (RTL) Collaborative Project}~\cite{rtlinux} maintains a kernel (based on the mainline Linux kernel) for real-time purposes. This patched kernel (known as the PREEMPT\_RT) ensures real-time behavior by making the scheduler fully preemptable. In this paper, we applied the PREEMPT\_RT patch 
%(version \texttt{4.9.47-rt37.patch.xz}) \hl{do we need to mention .patch.xz extension? - MH} 
on top of vanilla Raspbian (kernel version 4.9.48) to enable the real-time functionality.
To further enable the fully preemptive functionality from the PREEMPT\_RT patch, the $\mathtt{CONFIG\_PREEMPT\_RT\_FULL}$ flag was enabled during the kernel compilation phase. 
Furthermore, the system variable \texttt{/proc/sys/kernel/sched\_rt\_runtime\_us} was set to $-1$ to disable the throttling of the real-time scheduler.
This setting allowed the real-time tasks to use up the entire $100\%$ CPU utilization if required\footnote{This change in system variable settings was mainly configured for the purpose of  experimenting with the ideas of \papername only. For most real use-cases, users can keep this system variable untouched for more flexibility.}.
Also, the active core's \texttt{scaling\_governor } was set to ``\texttt{performance}'' mode to disable dynamic frequency scaling during the experiments.

\subsection{Vanilla EDF Scheduler}
%\todo[inline]{CY: What font style is better for the function name? texttt?}
%\hl{MH: I normally use mathtt since in IEEE format it gives you nice look :) -- also for RT\_PREEMPT let us use normal font (not italic) since previous papers use this way!}
%To implement the proposed algorithm and to easier evaluate its overhead, it is wise to start with an existing EDF scheduler.

Since Linux kernel version 3.14, an EDF implementation ($\mathtt{SCHED\_DEADLINE}$) is available in the kernel\cite{sched_dead}.
%That is, the operating system we select (\ie Raspbian with Linux kernel 4.9.48 and \textit{PREEMPT\_RT} patch 4.9.47-rt37) has support for $\mathtt{SCHED\_DEADLINE}$.
Since our PREEMPT\_RT patched kernel supports $\mathtt{SCHED\_DEADLINE}$, we used this as the baseline EDF implementation and extended the scheduler to implement the \papername protocol. %to realize the proposed randomization protocol. Besides the pure EDF scheduling algorithm, $\mathtt{SCHED\_DEADLINE}$ also implements a \textit{constant bandwidth server} (CBS) to handle the situation when a task overruns its declared WCET (either intentionally or unintentionally).
%However, since we only consider tasksets that are schedulable in this paper, CBS shall take no part in scheduling decisions as all tasks meet their deadlines.
%Therefore, we do not consider CBS in the following discussion.

% In a process\footnote{Since there is no distinction between processes and threads in the Linux kernel's scheduler, for the simplicity of the illustration, we use the term \textit{process}, \textit{thread} and \textit{task} interchangeably in the following context.}, 

In Linux the system call $\mathtt{sched\_setattr()}$ is invoked to configure the scheduling policy for a given process\footnote{Since there is no distinction between processes and threads in the Linux kernel's scheduler, for the simplicity of the illustration, we use the term \textit{process}, \textit{thread} and \textit{task} interchangeably in the following context.}.
By design, $\mathtt{SCHED\_DEADLINE}$ has the highest priority among all the supported scheduling policies (\eg $\mathtt{SCHED\_NORMAL}$, $\mathtt{SCHED\_FIFO}$ and $\mathtt{SCHED\_RR}$).
It's also worth noting that the Linux kernel maintains a separate run queue for $\mathtt{SCHED\_DEADLINE}$ (\ie $\mathtt{struct\ dl\_rq}$).
Therefore, it is possible to extend $\mathtt{SCHED\_DEADLINE}$ while keeping other scheduling policies untouched. Note that this vanilla EDF scheduler is also used as a base for comparison with the \papername protocol. The experimental results are presented in Section~\ref{subsec:linux}.
%\todo{Sibin: What was precomputed?}

\subsection{Implementation of \papername}

\subsubsection{Task/Job-specific Variables}
\label{sec:::tasl_variables}
%redf_budget for each job
The Linux kernel defines a structure, $\mathtt{struct\ sched\_dl\_entity}$, dedicated to $\mathtt{SCHED\_DEADLINE}$, to store task and job-related variables (both run-time and static variables).
They include typical EDF task parameters (\eg period, deadline and WCET).

To implement \papername we added two additional variables, named $\mathtt{reorder\_wcib}$ and $\mathtt{reorder\_rib}$, both $\mathtt{s64}$ (signed 64 bit integer) type variables, to store the WCIB for the task and to track the RIB for the task's active job at any given moment, respectively.
Each task's $\mathtt{reorder\_wcib}$ is initialized and updated when a new task is created.
The job-specific run-time variable, $\mathtt{reorder\_rib}$, is initialized to the precomputed $\mathtt{reorder\_wcib}$ every time when a new job arrives.
During run-time, the inversion budget was updated (\ie decreased by the elapsed time in the case of priority inversion) along with other $\mathtt{SCHED\_DEADLINE}$ run-time variables in the function $\mathtt{update\_curr\_dl()}$.
It is used to determine whether the inversion budget was consumed and a random selection of a job was allowed at a scheduling point.
%
\begin{comment}
To be precise, $\mathtt{redf\_budget}$ corresponds to the computation of $\Delta t$ at line 14 in Algorithm~\ref{alg:sched_rand}. 
\end{comment}

In our implementations we did not use any external libraries and only used the built-in kernel functions. The following listing shows a part of the existing variables as well as the newly added ones (the highlighted lines). %We omit the other variables for better readability.
Other variables added for the \papername protocol are shown in Appendix~\ref{sec::scheduler_variables}.
\vspace{-0.7\baselineskip}
% The source code
%\todo[inline]{Some comments exceed one line. -- MH: Fixed -- see whether makes sense}

\begin{center}
\begin{lstlisting}[language=C,escapechar=!,linebackgroundcolor={\ifnum\value{lstnumber}=6\color{shadecolor}\fi},]
struct sched_dl_entity {
/* task specific parameters */
	u64 dl_runtime;		// WCET
	u64 dl_deadline; 	// relative deadline
	u64 dl_period; 		// period 
	s64 reorder_wcib; // worst-case inversion budget
\end{lstlisting}
\vspace*{-0.6em}
\begin{lstlisting}[language=C,escapechar=!,linebackgroundcolor={\ifnum\value{lstnumber}=4\color{shadecolor}\fi},]
/* task instance (job) specific parameters */
	s64 runtime;			// remaining runtime 
	u64 deadline; 		// absolute deadline
	s64 reorder_rib; 	// remaining inversion budget
	....
/* Other variables are omitted for readability. */
};
\end{lstlisting}
\end{center}

\begin{comment}
\begin{center}
\begin{lstlisting}[language=C,escapechar=!,linebackgroundcolor={\ifnum\value{lstnumber}=6\color{shadecolor}\fi},]
struct sched_dl_entity {
/* task specific parameters */
	u64 dl_runtime;		// WCET
	u64 dl_deadline; 	// relative deadline
	u64 dl_period; 		// period 
	s64 redf_wcib; 		// worst-case inversion budget

/* task instance (job) specific parameters */
	s64 runtime;			// remaining runtime 
	u64 deadline; 		// absolute deadline
	s64 redf_rib; 		// remaining inversion budget
	....
/* Other variables are omitted for readability. */
};
\end{lstlisting}
\end{center}
\end{comment}

%!\colorbox{hl_code}{s64 redf\_budget;	// instance runtime budget}!
    
\vspace{-0.7\baselineskip}
\begin{comment}
\subsubsection{Taskset Variables}
%redf_taskset 
The computation of each task's $\mathtt{redf\_wcib}$ requires access to the taskset parameters.
For the sake of efficiency, we implemented the task pointers in a self-defined taskset structure. %stored in $\mathtt{dl\_rq}$.
A task pointer was stored when a process calls $\mathtt{sched\_setattr()}$ to switch its scheduling policy to $\mathtt{SCHED\_DEADLINE}$.
%D_min
Together with the task pointers, the upper bound of any busy-period length (\ie $\widehat{R}$, a taskset-specific value) was also stored in the taskset structure.
Recall that $\widehat{R}$ is needed when determining the limited priority inversion (line 3 in Algorithm~\ref{alg:sched_rand}).
Similar to the task pointers, we initialized and updated $D_{min}$ in the function $\mathtt{\_\_setscheduler()}$ whenever a process was joining $\mathtt{SCHED\_DEADLINE}$.
\end{comment}

\subsubsection{Task Selection Function}
\label{sec:task_selection_func}
%replace  pick_next_dl_entity with our function pick_rad_next_dl_entity
The \papername protocol was implemented as a function, named $\mathtt{pick\_rad\_next\_dl\_entity()}$, that selects a task and sets the next scheduling point based on the \papername algorithm.
It replaces the original $\mathtt{SCHED\_DEADLINE}$ function, $\mathtt{pick\_next\_dl\_entity()}$  (\ie one that picks the task that has the next absolute deadline from the run queue, \viz~the leftmost node in the scheduler's red-black tree).
This function is indirectly called by the main scheduler function $\mathtt{\_\_schedule()}$ when the next task for execution is needed.
%It's worth noting that these functions do not dequeue the picked task from the run queue.
%By design, a task is dequeued from the run queue when it

%\todo[inline]{We may need to add the sched\_dl\_entity as LaTeX listing figure and show some important fields, highlight our changes (perhaps sched\_deadline changes also)% -- say this struct:} 
% \begin{verbatim}
% struct sched_dl_entity {
% 	struct rb_node	rb_node;
% 	u64 dl_runtime;		
% 	u64 dl_deadline;	
% 	u64 dl_period;		
% 	u64 dl_bw;		
% .... and so on.. }
% \end{verbatim}

% CY: I think I get what you mean, don't bother to fix the error :) . I was planning to use listing to put some code there, but I'm still modifying my code. So I'll do that when they are final.

%Ok -- got that --MH!

\subsubsection{Randomization Function}
% Another option is prandom_u32().
We used the built-in random number generator in the kernel.
It supports the system call $\mathtt{get\_random\_bytes()}$ defined in \texttt{linux/random.h}.
It is used by the function $\mathtt{pick\_rad\_next\_dl\_entity()}$ to select a random task and a random execution interval for the next scheduling point as explained in Algorithm~\ref{alg:sched_rand}.
%using a shared entropy pool in the system.
%implemented in the file \texttt{drivers/char/random.c}

\subsubsection{Schedule Timer}
A high-resolution timer (\ie $\mathtt{struct\ hrtimer}$) was used to trigger the additional scheduling points introduced by the \papername protocol, as described in Algorithm~\ref{alg:sched_rand} (Line 22 and 23).
Since this timer is a scheduler-specific timer, it is stored in $\mathtt{dl\_rq}$, as $\mathtt{reorder\_pi\_timer}$.
It is worth noting that $\mathtt{hrtimer}$ is also used by $\mathtt{SCHED\_DEADLINE}$ to enforce the task periods.

\subsubsection{Idle Time Scheduling}
As introduced in Section~\ref{sec::idle_mode}, idle times are considered when the idle time scheduling scheme is deployed.
In our Linux kernel implementation, we utilized the native idle task maintained under the $\mathtt{SCHED\_IDLE}$ scheduler for this purpose.
The \papername protocol yields its scheduling opportunities (to other schedulers such as  $\mathtt{SCHED\_IDLE}$) if $\tau_I$, the idle task in the \papername protocol, is selected and running.
The subsequent scheduling point is 
%The duration of the idle time is 
enforced by $\mathtt{reorder\_pi\_timer}$.

% \section{\hl{Evaluation with Real-Time Linux Kernel on an Embedded Platform}}
\section{Evaluation}
%\label{sec:eval}
\label{subsec:linux}

%\input{Sections/eval_synthetic}

%\subsection{Results with Real-Time Linux Kernel on an Embedded Platform} 
%\label{subsec:linux}

In this section, we evaluate \papername using a prototype implemented on an embedded platform (\ie RPi3) running the real-time Linux kernel.
We mainly focus on {\em overheads for computing and selecting a task at each scheduling point}.
%and the proportion of random tasks selected at the scheduling points.
Recall that our implementation is based on the vanilla EDF scheduler, $\mathtt{SCHED\_DEADLINE}$, on Linux.
Therefore, we evaluate the overheads of the \papername protocol by comparing them with $\mathtt{SCHED\_DEADLINE}$.
The key observations from our performance evaluation results are summarized below.
\begin{itemize}

\item \papername works in practice on realistic embedded RTS and is able to meet the real-time guarantees.

\item The randomization logic adds minimal scheduling overhead in Linux kernel (Fig. \ref{fig:pick_next_task_cost}).
This overhead is arguably very small w.r.t. the task execution time (Fig. \ref{fig:cost_ratio}).
%Our prototype on Linux shows that \papername is more likely to select a random task when the system has more tasks (Fig. \ref{fig:random_task_proportion}).
\end{itemize}

\begin{comment}
% \subsection{Platform-based Evaluation}
%\subsection{Overhead Evaluation}
For evaluation on the real platform, we mainly focus on overheads for computing and selecting a task at each scheduling point.
%and the proportion of random tasks selected at the scheduling points.
Recall that our implementation is based on the vanilla EDF scheduler, $\mathtt{SCHED\_DEADLINE}$, on Linux.
Therefore, we evaluate the overheads of the \papername protocol by comparing them with $\mathtt{SCHED\_DEADLINE}$.
\end{comment}

% We measure the computation overhead inside the function $\mathtt{pick\_next\_task\_dl()}$.
% We want to point out for the reader that this function is different from the function $\mathtt{pick\_rad\_next\_dl\_entity()}$ or $\mathtt{pick\_next\_dl\_entity()}$ introduced in Section~\ref{sec:implementation}.
% It is a higher layer function that calls 

%WAIT: picking next task does not need to re-balance the dl\_rq as the leftmost node still has the closest absolute deadline. So the overhead may not be there.

\subsubsection{Experimental Setup}
We use the RPi3 platform as introduced in Section~\ref{sec:implementation}.
The operating system is patched and configured to enable the real-time capability, as shown in Table~\ref{tab:implementation_platform_summary}.
%The detailed configurations are shown in Table~\ref{tab:implementation_platform_summary}.
%
%As mentioned earlier, the vanilla EDF scheduler in Linux is tested as a baseline to understand the degree of the overhead.
To keep the vanilla EDF unpolluted from our implementation, we used two separately compiled kernels during the experiments.
In the vanilla EDF kernel, the scheduling functions remained untouched. 
Only the necessary code to benchmark the overhead were added.
%Note that the aforementioned configurations that enable the real-time capability were still applied on this kernel.
Note that PREEMPT\_RT real-time patch was still applied on this kernel.

%For these experiments on the real platform, $800$ synthetically generated tasksets were tested. The goal of the experiments was to evaluate the performance on synthetic workload on a real platform.
%For these experiments on the real platform, w
We used a mixture of  MiBench benchmark automotive programs~\cite{guthaus2001mibench} and synthetically generated tasks. 
The goal of the experiments was to evaluate the performance on both real and synthetic workloads on a real platform.
A total of $800$ tasksets were tested.
Each taskset was configured with the number of tasks from $1$ to $10$ ($10$ groups) and $50\%$ of the tasks are drawn from the MiBench programs (Table~\ref{tab:implementation_platform_summary}).
The utilization was set between the ranges $10\%$ and $90\%$ ($8$ utilization groups, $10$ tasksets per group) when generating the tasksets. 
Each task's period was randomly selected from the range $10$ ms and $5000$ ms. 
Taskset parameters were randomly generated using the taskset generator from the simulation (see Section \ref{subsec:sim_setup}).
The generated parameters (\eg the task's period and WCET) were multiples of $1ms$. 
In the experiments, the actual execution time performed by a synthetic task $\tau_i$ was limited to $\lfloor 0.8 \cdot C_i \rfloor$ (\ie $80\%$ of its WCET) to accommodate realistic task execution behaviors. 
%\hl{Can we use floor or Ceil function? (since we consider WCET as integer throughout)? -- MH}
%\hl{Time is stored as integers in Linux kernel - CY}
Both, vanilla EDF and \papername-based schedulers were tested with the same tasksets.

%To profile the context switches, we use the Linux command $\mathtt{perf}$, a powerful profiling tool, to record the performance counters during the experiments.
To profile the number of context switches, we directly recorded their occurrence in the scheduler.
We did not use external profiling tools 
%with performance counters 
(\eg $\mathtt{perf}$ \cite{perf}) because we only focus on the context switches that occur in 
%vanilla EDF and 
the $\mathtt{SCHED\_DEADLINE}$ scheduler (for both the vanilla EDF and the randomized EDF).
Using the profiling tool may include unnecessary context switch counts from other coexisting Linux schedulers. %before the tasks switch the scheduling policy to $\mathtt{SCHED\_DEADLINE}$.
To measure the execution time of the scheduling functions, the function $\mathtt{getnstimeofday()}$, defined in \texttt{linux/timekeeping.h}, was used.
For the experiments, we let each taskset run for $5$ seconds. The measurements and the scheduling trace were stored in the kernel log for further analysis.

\subsubsection{Results} \label{sec:linux_result}
\begin{figure}[!t]
\centering
%\advance\leftskip-0.8em
\includegraphics[width=0.88\linewidth,keepaspectratio]{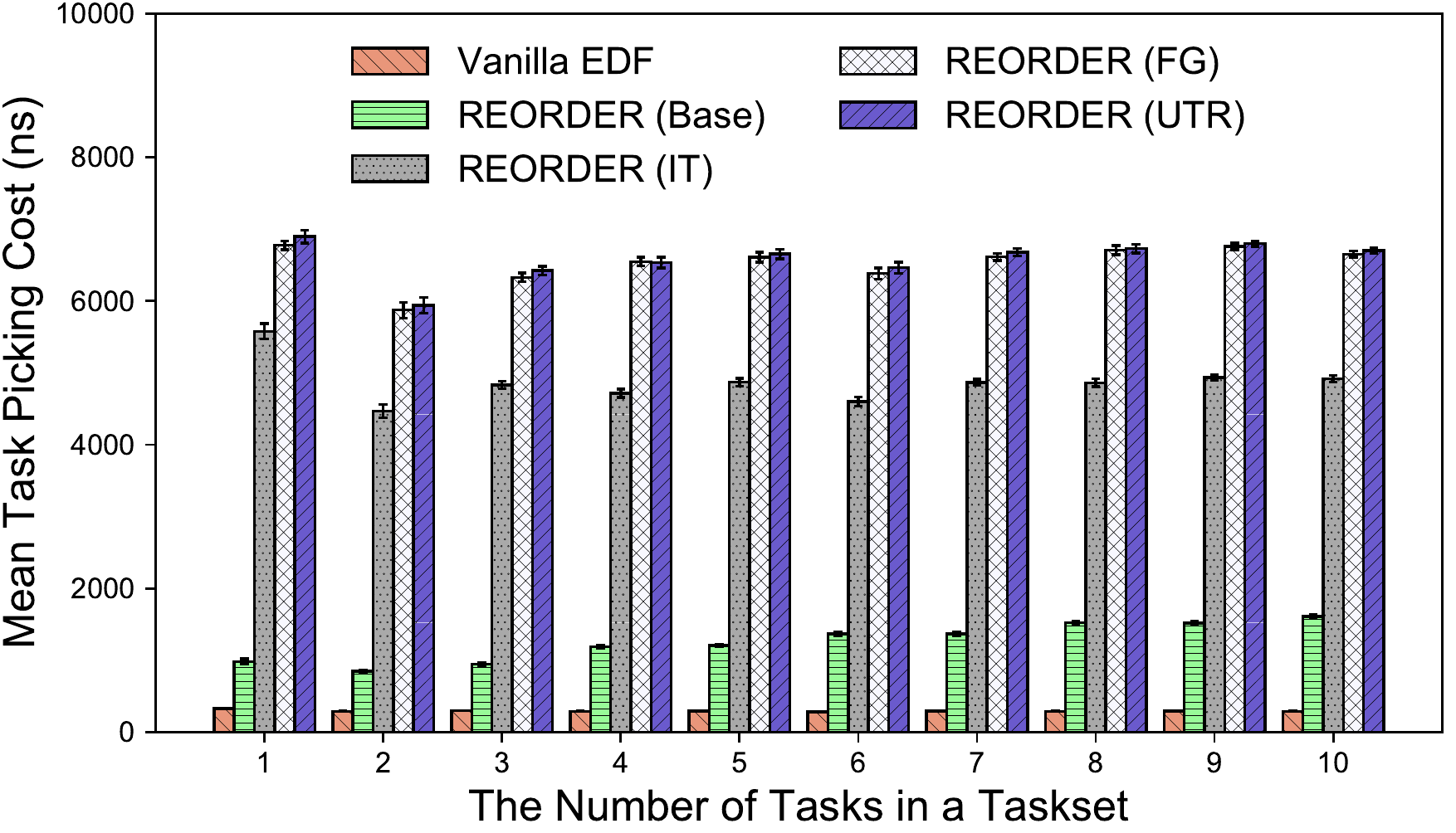}
\vspace{-0.5\baselineskip}
\caption{The execution time cost (in nanoseconds) for the scheduling functions of the vanilla EDF and the randomized EDF. 
%X-axis is the number of tasks in a taskset and Y-axis is the measured execution time in nanoseconds. 
The vanilla EDF bar represents the mean execution times processed by the function $\mathtt{pick\_next\_dl\_entity()}$ while the other three EDF bars present the mean execution times for $\mathtt{pick\_rad\_next\_dl\_entity()}$ that carries out the randomization algorithm.
} 
\label{fig:pick_next_task_cost}
\vspace{-1.5\baselineskip}
\end{figure}

We first examine the execution time overhead of the scheduling functions.
%in both vanilla EDF and randomized EDF scheduling functions.
%As mentioned in Section~\ref{sec:task_selection_func}, the main scheduling algorithms were implemented in the functions $\mathtt{pick\_next\_dl\_entity()}$ and $\mathtt{pick\_rad\_next\_dl\_entity()}$ for the vanilla EDF and the randomized EDF, respectively. 
As mentioned in Section~\ref{sec:task_selection_func}, the main algorithm for the \papername protocol was implemented in the function $\mathtt{pick\_rad\_next\_dl\_entity()}$.
This replaces the scheduling function $\mathtt{pick\_next\_dl\_entity()}$ in $\mathtt{SCHED\_DEADLINE}$ (vanilla EDF).
As this was the main change between the two schedulers, our test here was focused on measuring the execution time of $\mathtt{pick\_next\_dl\_entity()}$ (for vanilla EDF) and $\mathtt{pick\_rad\_next\_dl\_entity()}$ (for \papername) rather than the higher level scheduler function.
Fig.~\ref{fig:pick_next_task_cost} shows the results of this experiment.
%The corresponding raw data that includes the maximum/minimum values and the standard deviations is presented in Table~\ref{tab:cost_vanilla_edf} and Table~\ref{tab:cost_randomized_edf}.

From the figure, we can observe that the mean execution cost of $\mathtt{pick\_next\_dl\_entity()}$ for the vanilla EDF remains about the same across different taskset groups.
This result is expected because the vanilla EDF always selects the leftmost node from the Linux red-black tree (\ie run queue) that is independent to the number of tasks in a taskset and has complexity $O(1)$.
On the other hand, the mean execution cost of $\mathtt{pick\_rad\_next\_dl\_entity()}$ for the base randomization (without idle time randomization) is generally larger than the vanilla EDF mainly due to the $\mathtt{get\_random\_bytes()}$ calls (that takes an average $2531$ ns to generate a 64-bit random number) for the random task selections. 
When there is only one job in the run queue at a scheduling point, the base randomization scheme directly selects the job and omits the $\mathtt{get\_random\_bytes()}$ call.
%As a result, having fewer tasks in a taskset results in less frequent $\mathtt{get\_random\_bytes()}$ calls and hence has smaller mean scheduling overhead.
%These experiment results suggest that the mean overhead increases as the number of tasks in a taskset increases.
%
%In the worst case, the algorithm reaches the last step to pick a random task from the computed candidate list (that has complexity $O( |\mathcal{R}_\mathcal{Q}^t|)$ as discussed in Section~\ref{sec:algo_develop}).
%$O(n^2+n)$ mainly caused by line 2 and 3 in Algorithm~\ref{alg:sched_rand}. 
%\hl{@CY: The complexity analysis in Sec III-E and this one looks different? Can we make both consistent? I thought the complexity is order of $n$? -- MH}
%\todo[inline]{CY: Then I can just refer to your section. By the way, I think building the candidate list is $n^2$ (for each task it computes $\mathsf{IBF}$ which is a $O(n)$ function -- if it is not preciously computed and stored.)}
%\hl{I See what you mean since for DBF you need to iterate for all tasks}
%
%It's worth noting that the execution cost for the single-task taskset is higher than that of vanilla EDF.
%It is because of the computation of the minimum inversion deadline $m_{HP}^t$. Although one may argue that the implementation may be further optimized for the single-task taskset 
In the case of the idle time scheduling, \papername (IT), since the idle task is always considered in every scheduling point, the algorithm reaches the final step with a randomly selected task most of the time.
This leads to the overhead roughly corresponding to one $\mathtt{get\_random\_bytes()}$ call.
For the fine-grained switching with idle time randomization scheme and unused time reclamation (\ie \papername (FG/UTR)), the overhead remains at a higher level since, in the worst case, two $\mathtt{get\_random\_bytes()}$ calls are present for each scheduling point: one for the random task selection and the other for the random scheduling points.
This results in the scheduling overhead corresponding to the execution cost of two $\mathtt{get\_random\_bytes()}$ calls.
As a result, the overhead contributed by the other part of the algorithm that has complexity $O( |\mathcal{R}_\mathcal{Q}^t|)$ (as discussed in Section~\ref{sec:algo_develop})  is negligible compared to the randomization function.
%By measurement, $\mathtt{pick\_rad\_next\_dl\_entity()}$ takes an average $2531ns$ to generate a 64-bit random number which greatly dilutes the overhead influence from other factors (\eg the number of tasks in a taskset).

\begin{figure}[!t]
\centering
%\advance\leftskip-0.8em
\includegraphics[width=0.75\linewidth,keepaspectratio]{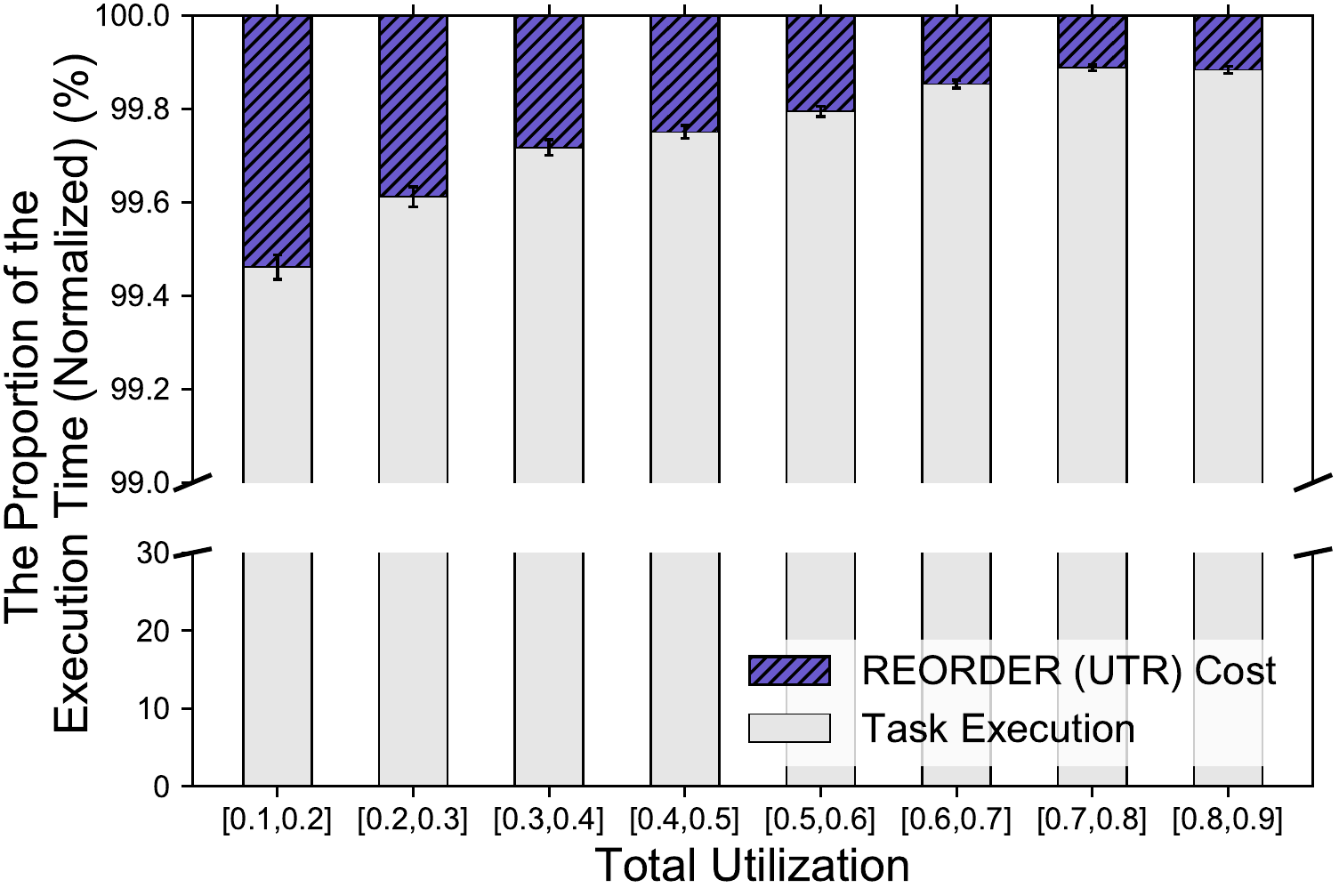}
\vspace{-0.5\baselineskip}
\caption{The proportion of the cost of the \papername protocol to the task execution times in the fine-grained switching with idle time randomization scheme. 
The randomization protocol overhead is provably inversely proportional to the taskset's total utilization. The upper part of the figure is scaled to $99\%-100\%$ for better readability.} 
\label{fig:cost_ratio}
\vspace{-1.5\baselineskip}
\end{figure}

% cost proportion
Next we examine the proportion of the scheduling overhead to the task's execution.
%timing impact of the scheduling overhead on the task's execution. 
We do this by comparing the cumulative time cost of the randomization protocol with the cumulative task execution times during the $5$ second test duration for each taskset.
Here, we consider the fine-grained switching with idle time randomization and unused time reclamation scheme (\papername (UTR)) as it has the largest overhead among all possible schemes.
Fig.~\ref{fig:cost_ratio} shows the mean proportion of the cost of the \papername (UTR) protocol to the task execution times with varying total utilization.
The results indicate that the overhead of the \papername protocol is
inversely proportional to the taskset's total utilization.
Since a taskset with higher utilization spends more of its time executing actual RT jobs, it dilutes the influence from the overhead.
The utilization group $[0.8, 0.9]$ has an average of $0.12\%$ overhead while it is 
$0.54\%$ for the $[0.1, 0.2]$ utilization group.
Considering there is typically an overestimation in the range $5\%$ and $15\%$ for task WCET calculations \cite{Wilhelm:2008:WEP}, %we argue that 
the overhead of the \papername protocol is negligible for most RTS.

\section{Discussion}

Although we only focused on the fact that \papername can reduce the predictability of conventional dynamic priority scheduler, this idea improves the security posture of future RTS in a more fundamental way. For any scheduling policy, one can infer the amount of information \textit{leaked} from the system. This information, for instance, will be useful for the engineers to analyze the potential \textit{vulnerability} (associated with timing inference attacks) of the given system. 

Consider a schedule $\mathcal{S}$ that is output from the randomization protocol (referred to as \textit{ground-truth process}) and let $\mathcal{S}^\prime$ be the attackers (potentially semi-correct) observation about the schedule (noted as \textit{observation}). We can define the information leakage as the amount of uncertainty (of the adversary) as follows: the uncertainty about the ground-truth process minus the attackers uncertainty (about the true schedule) after receiving the (fuzzy) observation (\ie the amount of the reduction of uncertainty due to receiving the observation). One can then use \textit{mutual information} \cite[Ch. 2]{information_theory_book} (\eg $\underset{\tilde{u}}{\sup}~ I(\mathcal{S};\mathcal{S}^\prime)$ where $\tilde{u}$ is the possible decoding strategies that an adversary can use and $I(\mathcal{S};\mathcal{S}^\prime) = H(\mathcal{S})-H(\mathcal{S}|\mathcal{S}^\prime)$ is the mutual information) between the ground-truth and the observation %or the \textit{min-entropy based leakage}, 
as a measure of leakage. A high dependency between the ground-truth  and the observation leads to a high information leakage. This implies that the adversary can have a good estimation of the ground-truth. The frameworks developed in this work aims to increase the randomness of the output of the scheduler and reduce the dependency between $\mathcal{S}$ and the $\mathcal{S}^\prime$. This is because, for the randomized scheduler, there are more true schedules that are consistent with a given observation. We highlight that defining the exact relationship between the produced randomness and the leakage of the system will require further study.
%a rigorous study of leakage for our proposed approach will d. 
We intend to explore this aspect in future work.

While \papername reduces the chances of the success of timing inference attacks (and hence improves the security), it is \textit{not} free from trade-offs. For instance, as we observe in Fig. \ref{fig:pick_next_task_cost} and \ref{fig:cost_ratio}, the randomization logic adds extra overheads to the scheduler. %Besides, randomization also increases the number of context switches (see Fig. \ref{fig:context_switches}). 
In this work we did not attempt to derive any analytic upper-bound on the number of context switches and leave this for future work.

%. Our future work will focus on designing a \textit{tunable framework} where the system engineers can control parameters depending on application requirements (\eg for a given taskset $\Gamma$, limit the number of context switches up to $\varepsilon_{cs}$ with a trade-off in $\Delta \%$ reduced randomness)  that provide best trade-off between system performance and security.

Note that it may be possible that some (heavily utilized) tasksets can \textit{not} be randomized and in that case both EDF and \papername output the same schedule. For instance, let us consider the taskset  $\Gamma_{ex3}= \lbrace \tau_1, \tau_2, \tau_3, \tau_4 \rbrace$ with the following parameters: $C_1 = 1, C_2 = 3, C_3 = 2, C_4 = 4$ and $T_1 = 5, T_2 = 8, T_3 = 9, T_4 = 20$ (with $T_i=D_i, 1 \leq i \leq 4$). The taskset is schedulable by EDF since $\sum\limits_{1 \leq i \leq 4} \frac{C_i}{T_i} = 0.997 < 1$. However, in this case the budgets (\eg WCIB) are always negative for all the tasks, \eg $V_1 = -2, V_2 = -1, V_3 = -4, V_4 = -4$. Therefore, at each scheduling point \textit{all} the low-priority jobs will be excluded from priority inversion and only the shortest deadline job will be selected -- \ie the same schedule as EDF. 

%and thus the system may be exposed to similar vulnerabilities as EDF.

\section{Related Work}
\label{related}
% Security in RTS

% Security in real-time schedule
% Attack:
% ScheduLeak

% Protection:
% Sibin's no-leak paper.

% Randomization techniques

% Schedule Randomization

% TaskShuffler~\cite{2016:taskshuffler}

% "Improving Security for Time-Triggered Real-Time Systems against Timing Inference Based Attacks by Schedule Obfuscation" 

% \hl{TODO: Talk about ECRTS paper?}

Kr{\"u}ger \etal~\cite{volp_TT_randomization} proposed a combined online/offline randomization scheme to reduce determinisms for time-triggered (TT) systems where tasks are executed based on a pre-computed, offline, slot-based schedule. The scheduling paradigms for TT systems are different than dynamic priority RTS.
%and thus those schemes can not be applied directly for a RTS scheduled by EDF. 
The closest line of work is TaskShuffler~\cite{taskshuffler} where authors proposed to randomize task schedules for fixed-priority (\eg RM) systems. However the methods developed in both of the above are not directly applicable for dynamic priority systems. Unlike fixed priority systems, obfuscating schedules for EDF scheduling is not straightforward due to run-time changes to task priorities.  Besides, as we describe in Appendix~\ref{subsec:entropy_limitation}, the calculation of schedule entropy in prior work does not correctly capture the randomness  for all scenarios.  Prior work also assumes all the jobs of the tasks always execute with WCET and hence may not be practical for real applications.

Zimmer \etal~\cite{time_zimmer} propose the mechanisms to detect the execution of unauthorized
instructions that leverages the information obtained by static timing analysis. 
An architectural approach that aims to create hardware/software mechanisms to detect anomalies is studied by Yoon \etal~\cite{mohan_s3a}.
Threats to covert timing channels for RTS has been addressed in prior research for fixed-priority systems \cite{leak2}. A scheduler-level modification is proposed in literature \cite{leak3} that alters thread blocks (that may leak information) to the idle thread – the aim is to avoid the exploitation of timing channels while achieving real-time guarantees. The authors also developed locking protocols for preventing covert channels \cite{leak1}.

Issues regarding information leakage through storage timing channels (\eg caches) in RTS, with different security levels, has been studied \cite{sg1, sg2} and further generalized \cite{sibin_RT_security_journal}. The authors proposed a modification to the fixed-priority scheduling algorithm and introduced a state cleanup mechanism to mitigate information leakage through shared resources. However, this leakage prevention comes at a cost of reduced schedulability and is focused on fixed-priority systems. 
Besides, they may not be completely effective against timing inference attacks that focus on deterministic scheduling behaviors. \papername works to break this inherent predictability of real-time scheduling by introducing randomness.

Bao \etal \cite{thermal_profile_ccs} model the behavior
of the attacker and introduce a scheduling  algorithm for a system with aperiodic tasks that have \textit{soft} deadlines. They provide a trade-off between side-channel information
leakage and the number of deadline misses. To the best of our knowledge \papername is the first work that focuses on obfuscating schedule timing information for dynamic priority RTS with hard deadlines.

\section{Conclusion}
\label{sec::concl}

Malicious attacks on systems with safety-critical real-time requirements could be catastrophic since the attackers can destabilize the system by inferring the  critical task execution patterns. In this work we focus on a widely used optimal real-time scheduling policy and make progress towards developing a solution for timing side-channel attacks. By using the approaches developed in this work (along with our open-source Linux kernel implementation) engineers of the systems can now have enough flexibility, as part of their design, to secure such safety-critical systems. While our initial findings are promising, we believe this is only a start towards developing a unified secure real-time framework in general.

%\newpage
\appendix

%\section{Appendix Section I}

% \input{Sections/th_proof}

% \input{Sections/algorithm}

\subsection{Calculation of an Upper Bound of the Response Time} \label{sec:wcrt_calculation}

Under EDF, the response time calculation involves computing the \textit{busy-period}\footnote{A busy-period \cite{busy_period} of $\tau_i$ is the interval $[t_0, t]$ within which jobs with priority higher or equal than $\tau_i$ are processed throughout $[t_0, t]$ but no jobs with priority higher or equal than $\tau_i$ are processed in $t_0 - \epsilon, t_0)$ or $(t, t + \epsilon)$ for a sufficiently small $\epsilon$.} of a task's instance with deadline less than or equal to that instance \cite{edf_wcrt_2}. Real-time theory uses the notion of \textit{interference}, \eg the amount of time a ready job of $\tau_i$ is blocked due to the execution of other higher priority jobs.
To calculate the WCIB of a task, we measure the worst-case interference from its higher priority jobs.  Note that with arbitrary priority
inversions, any job could be delayed because of chain reactions, \ie some low priority jobs in $lp(\tau_i)$, delay the higher priority jobs (\eg $\tau_j, \forall d_j < d_i$), that in turn delay $\tau_i$ -- hence $\tau_i$ may need more than its WCRT as calculated by the response time analysis \cite{res_time_rts,edf_wcrt_1}. This phenomenon is known as back-to-back hit \cite{res_time_rts} and can be addressed by considering an extra instance of higher priority jobs. Therefore, without any assumptions on the execution patterns of $lp(\tau_i)$, for a given release time $t=a$ we can calculate the upper bound of interference \cite{edf_wcrt_1, edf_wcrt_2, res_time_rts} experienced by $\tau_i$ as follows:
\begin{equation} \label{eq:edf_intf}
I_i(a) = \hspace*{-1em} \sum_{j \neq i, D_j \leq a + D_i} \hspace*{-1em} \min \left\lbrace \left\lceil \tfrac{D_i}{T_j} \right\rceil + 1 , 1 + \left\lfloor \tfrac{a + D_i - D_j}{T_j} \right\rfloor + 1  \right\rbrace C_j.
\end{equation}
Note that the extra execution times (\eg $+1$ in Eq.~(\ref{eq:edf_intf})) are added in the interference calculation to prevent the effects of back-to-back hit from higher priority jobs. 
For a given release time $a$, the response time of $\tau_i$ \cite{edf_wcrt_1, edf_wcrt_2} (relative to $a$) is given by: $R_i(a) = \max \left\lbrace C_i, W_i(a) - a \right\rbrace$ where $W_i(a)$ denotes the workload of $\tau_i$ and calculated by $W_i(a) = \left( \left\lfloor \frac{a}{T_i} \right\rfloor +1 \right)C_i + I_i(a)$. 
%and $I_i(a)$ is calculated by Eq. (\ref{eq:edf_intf}). 
Finally we can compute the upper bound of WCRT of $\tau_i$ as follows:
%\begin{equation}\label{eq:upper_bound_wcrt_i}
$
R_i = \max \left\lbrace R_i(a) \right\rbrace,~ 0 \leq a < \widehat{R} - C_i
$
%\end{equation}
where  $\widehat{R}$ is calculated by an iterative fixed point search, that is $\widehat{R} = r^{(k+1)} = r^{(k)}$ for some iteration $k$ where $r^{(\cdot)}$ is the upper bound of any busy-period length. We can calculate this upper bound using the following recurrence relation:
%\begin{eqnarray}
$
r^{(0)} = \sum\limits_{\tau_i \in \Gamma} C_i,~ 
%\quad  %\\
r^{(k+1)}  = \sum\limits_{\tau_i \in \Gamma} \left\lceil \frac{r^{(k)}}{T_i} \right\rceil C_i
$.
%\end{eqnarray}
This sequence $r^{(k)}$ converges to $\widehat{R}$ in a finite number of steps if we assume that the taskset is schedulable (\ie $\sum\limits_{\tau_i \in \Gamma}\frac{C_i}{T_i} \leq 1$) \cite{edf_wcrt_2}.

\subsection{Limitations of Existing Entropy Calculation Approach} \label{subsec:entropy_limitation}

% \hl{move this section to appendix}

In order to evaluate the performance of a randomized scheduler, we need to have a measure of the randomness of the output of the scheduler. 
For a taskset with hyperperiod of length $L$, define the $L$ dimensional random vector $\mathcal{S}^k=\left[S_1^k,\cdots,S_L^k\right]$ representing the schedule of hyperperiod $k$, where the random variable $S_t^k\in\{\tau_0,...,\tau_n\}$ denotes the task (including the idle task) scheduled at the $t$-th slot of hyperperiod $k$.
Note that the random vectors $\mathcal{S}^k$ for different values of $k$ are  independent and identically distributed (i.i.d.) random variables. Therefore the average randomness of the whole output is equal to the randomness in a single hyperperiod. 
 
In the past~\cite{taskshuffler}, researchers defined the entropy of the schedule $\mathcal{S}^k$ using Shannon entropy \cite[Ch. 2]{information_theory_book} as a measure of the randomness, \ie
% \begin{align}
% \label{eq:entropy}
$H(\mathcal{S}^k) = - \hspace*{-2em} \sum\limits_{s_1^k\in\{\tau_0,\cdots,\tau_n\}} 
\hspace*{-0.5em} 
\cdots 
\hspace*{-0.5em}
\sum\limits_{s_L^k\in\{\tau_0,\cdots,\tau_n\}} \hspace*{-0.5em}  \mathbb{P}(\mathcal{S}^k=[s_1,\cdots,s_L])
\times 
\log_2 \mathbb{P}(\mathcal{S}^k=[s_1^k,\cdots,s_L^k]) 
%\end{align}
$
with the assumption that $0 \times \log_2 0=0$.
% %\begin{equation}
% \begin{align}
% \label{eq:entropy}
% H(\mathcal{S}^k) = - \hspace*{-2em} \sum\limits_{s_1^k\in\{\tau_0,\cdots,\tau_n\}} 
% \hspace*{-0.5em} 
% \cdots 
% \hspace*{-0.5em}
% \sum\limits_{s_L^k\in\{\tau_0,\cdots,\tau_n\}} \hspace*{-0.5em} & \mathbb{P}(\mathcal{S}^k=[s_1,\cdots,s_L])
% \times \nonumber \\
% \log_2 \mathbb{P}(\mathcal{S}^k=[s_1^k,\cdots,s_L^k]) \hspace*{-5em}
% \end{align}
% %\end{equation}
There are two major issues in calculating the schedule entropy using the above method. 

\textit{First}, in order to obtain $H(\mathcal{S}^k)$, we need to calculate the distribution $\mathbb{P}({\mathcal{S}^k})$ -- calculating this distribution has exponential complexity and is not computationally tractable in practice. Also, estimating this distribution requires a very high number of samples. To address this problem, we proposed \cite{taskshuffler} to use the sum of the entropy of random variables $S_t^k$, $t \in\{1,..,L\}$ as the measure of randomness (referred to as upper-approximated schedule entropy):
%\begin{equation}\label{eq:sumentropy}
$
\widetilde{H}(\mathcal{S}^k)=\sum_{t=1}^L H(S_t^k),
$
%\end{equation}
where $H(S_t^k) = - \hspace*{-1em} \sum\limits_{s_t^k\in\{\tau_0,\cdots,\tau_n\}} \hspace*{-1em} \mathbb{P}(S_t^k=s_t^k) \log_2 \mathbb{P}(S_t^k=s_t^k)$. Note that for $\pi=0$, choosing $m=1$ gives us $\widehat{H}(\mathcal{S}^k, m,\pi,K)=\widetilde{H}(\mathcal{S}^k)$ and choosing $m=L$ outputs $\widehat{H}(\mathcal{S}^k, m,\pi,K)\rightarrow{H}(\mathcal{S}^k)$, as $K\rightarrow\infty$. The main limitations of upper approximated schedule entropy $\widetilde{H}(\mathcal{S}^k)$ is that it completely ignores the regularities that exist in $\mathcal{S}^k$ due to the dependencies among random variables. For instance, suppose a taskset contains two tasks: $\Gamma = \lbrace \tau_1, \tau_2 \rbrace$ and a schedule for first $5$ slots in $2$ individual hyperperiods is as follows: $\mathcal{S}_1\in\{(\tau_1,\tau_2,\tau_1,\tau_2,\tau_1),(\tau_2,\tau_1,\tau_2,\tau_1,\tau_2)\}$, assuming each vector with equal probability. Let us consider another schedule $\mathcal{S}_2$ that has all possible $2^5$ vectors of $\tau_1$ and $\tau_2$ of length $5$ with equal probability. Then $\widetilde{H}(\mathcal{S}_1)=\widetilde{H}(\mathcal{S}_2)$ even though the randomness of $\mathcal{S}_2$ is much higher (\ie $H(\mathcal{S}_1)=1$ while $H(\mathcal{S}_2)=5$). Therefore, $\widetilde{H}(\mathcal{S}^k)$ cannot capture the randomness correctly.

\textit{Second}, consider an instance where many of the schedules produced in different hyperperiods have very similar patterns in the first few slots and different patterns in the latter slots (or vice versa). In such cases $H(\mathcal{S}^k)$ cannot capture the similarities and considers the observed hyperperiods as distinct ones -- this leads us to search for dissimilarities in \textit{intervals} smaller than the whole length of the hyperperiod. In what follows we propose an entropy measure to capture the randomness of a schedule using the concept of \textit{limited size intervals} that resolves both the aforementioned issues and provides a better way to quantitatively compute randomness. 

%Note that for $\pi=0$, choosing $m=1$ gives us $\widehat{H}(\mathcal{S}^k, m,\pi,K)=\widetilde{H}(\mathcal{S}^k)$ and choosing $m=L$ outputs $\widehat{H}(\mathcal{S}^k, m,\pi,K)\rightarrow{H}(\mathcal{S}^k)$, as $K\rightarrow\infty$, where $\widetilde{H}(\mathcal{S}^k)$ and $H(\mathcal{S}^k)$ are defined in Eqs. \eqref{eq:sumentropy} and \eqref{eq:entropy}, respectively.

\subsection{Comparison With True and Approximate Entropy} \label{appsec:tru_apen_cor}

Recall that obtaining the true entropy (\eg $H(\cdot)$) is not feasible in practice since it has an asymptotic complexity. Therefore, we compare approximate entropy (\eg $\widehat{H}(\cdot)$ with $H(\cdot)$ by measuring the correlation observed from small tasksets. We generate tasksets that have $[3, 5]$ tasks with $T_i \in \lbrace 2, 4,5,10,20 \rbrace$ where the task utilizations and WCET are generated using methods from  Section \ref{subsec:sim_setup}. Each taskset has a common hyperperiod $L=20$ (allowing us to evaluate enough schedules for a reasonable time). For each taskset we observe the schedule for $K=1500$ hyperperiods and estimate the true entropy. Given a fixed taskset, generating unique schedules (\eg $K \rightarrow \infty$) leads to actual entropy $H(\cdot)$ since more tasks appear at each slot.
For approximate entropy $\widehat{H}(\cdot)$ we set interval length
$m=\lceil 0.35 L \rceil$ and dissimilarity threshold $\pi=0.1 L$ by trial-and-error and measure the correlation.

\begin{figure}[h]
\centering
%\advance\leftskip-0.8em
\vspace{-0.5\baselineskip}
\includegraphics[width=2.5in]{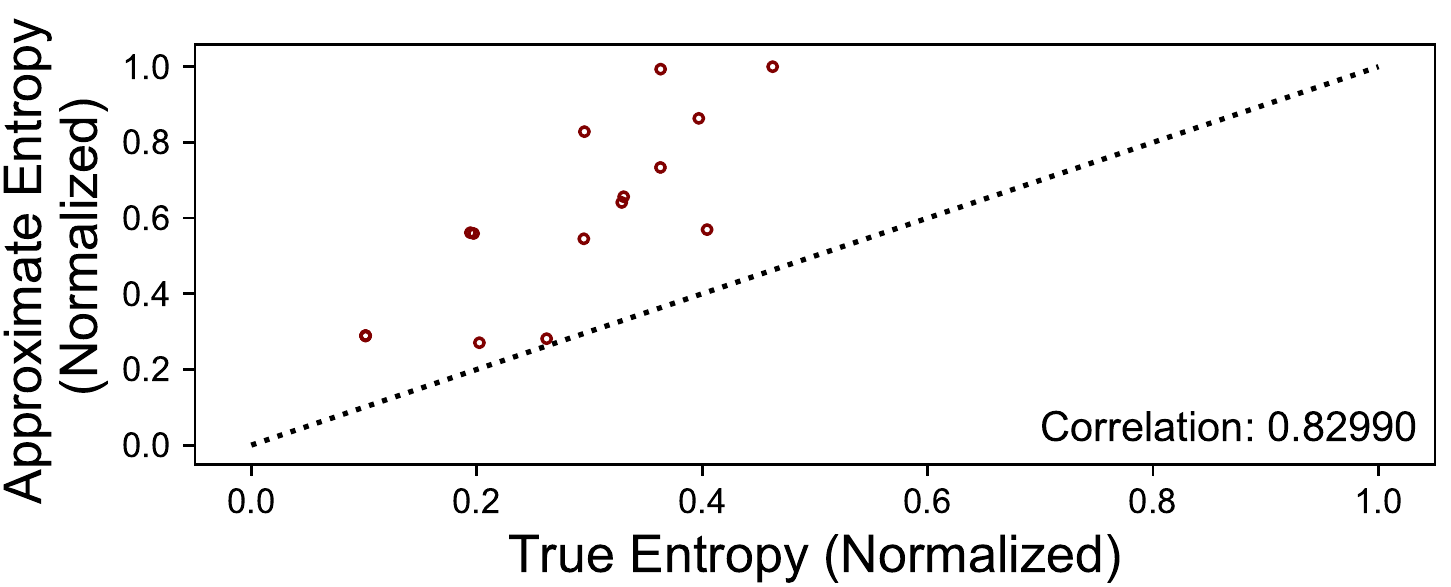}
\vspace{-0.5\baselineskip}
\caption{The correlation between true and approximate entropy (the values are normalized to $[0,1]$).} 
\vspace{-0.7\baselineskip}
\label{fig:tru_en_cor}
\end{figure}

The true and approximate entropy do not depend on the length of the hyperperiod -- instead, the approximation error (as can be seen from Fig. 7) is due to the assumption of independence between intervals. While we observe that the correlation between true and approximate entropy is relatively high (\eg $\approx\!\!0.82$) the approximated schedule entropy, $\widehat{H}(\cdot)$ should be used to compare the \textit{relative randomness} of two schedules (that is also the focus of our evaluation). 

%we measure the correlation between the true and approximate entropy (\eg $\widehat{H}(\cdot$) for a small taskset.

%\todo[inline]{Sibin: Some more discussion on how to pick $r$.}

%=============subsection==============
\subsection{\papername Variables in Real-time Linux Implementation}
\label{sec::scheduler_variables}
The implementation of the \papername protocol on the Linux kernel modifies four files:
\begin{itemize}
\item $\texttt{include/linux/sched.h}$ (task/job-specific variables introduced in Section~\ref{sec:::tasl_variables}).
\item $\texttt{kernel/sched/sched.h}$ (scheduler-specific variables, as presented below).
\item $\texttt{kernel/sched/core.c}$ (scheduling functions that govern all schedulers in the kernel).
\item $\texttt{kernel/sched/deadline.c}$ (scheduling functions for $\mathtt{SCHED\_DEADLINE}$ -- the main \papername algorithms were implemented here).
\end{itemize}

Besides the task-specific variables introduced in Section~\ref{sec:::tasl_variables}, there are scheduler-specific variables declared and used in our implementation, as shown in the listing below.
\vspace{-0.7\baselineskip}
\begin{center}
\begin{lstlisting}[language=C++,escapechar=!,linebackgroundcolor={\ifnum\value{lstnumber}=100\color{shadecolor}\fi},]
struct dl_rq {
	....
/* scheduler specific parameters */
	struct hrtimer reorder_pi_timer; // schedule timer 
	u64 reorder_pi_timer_start_time; // timer start
	bool reorder_idle_time_acting;   // idle status
  // scheme (Base, IT, FT, UTR)
	enum reorder_scheduling_mode reorder_mode; 
	....
};
\end{lstlisting}
\end{center}

% /* other variables are omitted for readability. */
\vspace{-0.7\baselineskip}
The variable $\mathtt{reorder\_mode}$ is used to determine the randomization scheme to be used in the scheduler.
The enumeration for the scheme options are defined in the same source file ($\texttt{kernel/sched/sched.h}$) and shown in the following listing.
\vspace{-0.7\baselineskip}
\begin{center}
\vspace{-0.5\baselineskip}
\begin{lstlisting}[language=C,escapechar=!,linebackgroundcolor={\ifnum\value{lstnumber}=100\color{shadecolor}\fi},]
enum reorder_scheduling_mode {
	REORDER_NORMAL,      // task only randomization
	REORDER_IDLE_TIME,   // + idle time scheduling
	REORDER_FINE_GRAINED,// + fine-grained switching
	REORDER_RECLAMATION  // + unused time reclamation
};
\end{lstlisting}
\vspace{-0.5\baselineskip}
\end{center}

%\newpage
\bibliographystyle{IEEEtran}
% \bibliographystyle{ACM-Reference-Format}
%\bibliography{ref,ref_monowar,sibin.security}
\bibliography{main}

\end{document}